%% file: main.tex
\documentclass[sigconf]{acmart}
\usepackage{amsmath}
\usepackage{url}
\usepackage{subfigure}
\usepackage{graphicx}
\usepackage{mdframed}
\usepackage{enumitem}
\usepackage[linesnumbered,ruled,vlined,lined,boxed,commentsnumbered]{algorithm2e}
\newmdenv[
    backgroundcolor=gray!20,
    linecolor=black,
    linewidth=1pt,
    roundcorner=5pt,
    frametitle={Prompt},
    frametitlefont=\bfseries,
    frametitlebackgroundcolor=gray!30,
]{promptbox}
\usepackage{balance}
\usepackage{hyperref}
\hypersetup{hidelinks,
	colorlinks=true,
	allcolors=black,
	pdfstartview=Fit,
	breaklinks=true}

\copyrightyear{2025}
\acmYear{2025}
\setcopyright{cc}
\setcctype{by}
\acmConference[MM '25]{Proceedings of the 33rd ACM International Conference on Multimedia}{October 27--31, 2025}{Dublin, Ireland}
\acmBooktitle{Proceedings of the 33rd ACM International Conference on Multimedia (MM '25), October 27--31, 2025, Dublin, Ireland}\acmDOI{10.1145/3746027.3755257}
\acmISBN{979-8-4007-2035-2/2025/10}
\begin{document}

\title{Beyond Interpretability: Exploring the Comprehensibility of Adaptive Video Streaming through Large Language Models
}

\author{Lianchen Jia}
\affiliation{%
  \institution{Department of Computer Science and Technology, Tsinghua University}
  \city{Beijing}
  \country{China}
}
\email{jlc21@mails.tsinghua.edu.cn}

\author{Chaoyang Li}
\affiliation{%
  \institution{Department of Computer Science and Technology, Tsinghua University}
  \city{Beijing}
  \country{China}
}
\email{1161067272@qq.com}
\author{Ziqi Yuan}
\affiliation{%
  \institution{Department of Computer Science and Technology, Tsinghua University}
  \city{Beijing}
  \country{China}
}
\email{yzq21@mails.tsinghua.edu.cn}
\author{Jiahui Chen}
\affiliation{%
  \institution{Department of Computer Science and Technology, Tsinghua University}
  \city{Beijing}
  \country{China}
}
\email{chenjiah22@mails.tsinghua.edu.cn}
\author{Tianchi Huang}
\affiliation{%
  \institution{Department of Computer Science and Technology, Tsinghua University}
  \city{Beijing}
  \country{China}
}
\email{mythkastgod@gmail.com}
\author{Jiangchuan Liu}
\affiliation{%
  \institution{School of Computing Science, Simon Fraser University}
  \city{Burnaby}
  \country{Canada}
}
\email{jcliu@cs.sfu.ca}
\author{Lifeng Sun}
\authornote{Lifeng Sun is the corresponding author.}
\affiliation{%
  \institution{Department of Computer Science and Technology, Tsinghua University}
  \institution{ Beijing National Research Center for Information Science and Technology
}
  \city{Beijing}
  \country{China}
}
\email{sunlf@tsinghua.edu.cn}
\renewcommand{\shortauthors}{Lianchen Jia et al.}


\input{abstract}

\maketitle

\input{intro}

\input{MOTIVATION}

\input{system}
\input{evaluation}
\input{discuss}
\input{conclude}


\newpage
\balance
\bibliographystyle{ACM-Reference-Format}
\bibliography{ref}
\newpage
\input{app}
\end{document}

%% file: abstract.tex
\begin{abstract}
Over the past decade, adaptive video streaming technology has witnessed significant advancements, particularly driven by the rapid evolution of deep learning techniques. However, the black-box nature of deep learning algorithms presents challenges for developers in understanding decision-making processes and optimizing for specific application scenarios. Although existing research has enhanced algorithm interpretability through decision tree conversion, interpretability does not directly equate to developers' subjective comprehensibility. To address this challenge, we introduce \texttt{ComTree}, the first bitrate adaptation algorithm generation framework that considers comprehensibility. The framework initially generates the complete set of decision trees that meet performance requirements, then leverages large language models to evaluate these trees for developer comprehensibility, ultimately selecting solutions that best facilitate human understanding and enhancement. Experimental results demonstrate that \texttt{ComTree} significantly improves comprehensibility while maintaining competitive performance, showing potential for further advancement. The source code is available at \url{https://github.com/thu-media/ComTree}.
\end{abstract}

%% file: intro.tex
\section{INTRODUCTION}
With the advancement of networking technologies~\cite{narayanan20215g,wifi6} and the growing desire for content creation~\cite{pires2015ugc}, video has become an indispensable part of people's daily lives. According to the 2024 Global Internet Phenomena Report~\cite{sandvine24}, video applications are the primary contributors to downstream traffic, accounting for 39\% of fixed network traffic and 31\% of mobile network traffic. Within content categories, on-demand streaming represents 54\% of downstream traffic in fixed networks and 57\% in mobile networks, making it the largest subcategory. Furthermore, according to the 2025 Global Internet Phenomena Report~\cite{sandvine25}, the top three major video applications have experienced traffic growth of 15\%--51\% in fixed networks and 7\%--38\% in mobile networks compared to the previous year. In video streaming technology, Adaptive Bitrate (ABR) has emerged as the mainstream approach for enhancing the Quality of Experience (QoE) of network video traffic.

ABR has been evolving for more than a decade~\cite{sani2017abrsurvey}. These algorithms select the appropriate bitrate level for the next segment based on the current buffer size and historical bandwidth conditions. ABR algorithms aim to enhance the user's QoE by improving video quality and playback smoothness while reducing stalling events. 

Traditional heuristic algorithms optimize the bitrate adaptation process by modeling the playback process~\cite{bba,yin2015mpc,spiteri2020bola}. In recent years, deep learning algorithms have shown promising results in bitrate adaptation~\cite{mao2017pensieve,huang2019comyco,jia2023rdladder,jia2024dancing,wu2024netllm}. These algorithms have the potential to automatically adapt to changes in the environment. However, when deploying these algorithms in production environments, the black-box optimization nature of deep learning brings difficulties for developers to understand. Improving the interpretability of learning-based ABR models has become a hot topic of interest. These works either understand the importance of black-box model states through post-hoc analysis~\cite{dethise2019cracking} or transform the black-box models into decision trees for deployment~\cite{meng2019pitree,gruner2020reconstructing}.

However, for ABR algorithms, merely transforming them into white-box algorithms with interpretability for each decision step is insufficient. A practical challenge in production environments is whether developers can comprehend and further optimize the algorithm to adapt to new network conditions - this refers to the algorithm's comprehensibility~\cite{huysmans2011comprehensibility,freitas2014comprehensible,martens2011comprehensibilit,freitas2006we,pazzani2000knowledge}. Although previous approaches using greedy algorithms to generate decision trees provide reasons for each decision step, complex decision logic~\cite{gruner2020reconstructing,meng2019pitree} may still pose comprehension challenges for developers~(e.g., tree depth, node selection/organization). This complexity creates barriers when developers attempt to adjust and optimize the decision tree. Therefore, our objective extends beyond mere interpretability, aiming to enhance algorithm comprehensibility and ultimately maximize the potential for developer-driven improvements.

While enhancing the comprehensibility of ABR shows promising potential, two significant challenges emerge. The first challenge lies in improving algorithm comprehensibility without substantially compromising performance. Current decision tree generation algorithms prioritize performance through greedy node selection until growth stops or node limits are reached~\cite{loh2011cart}. However, merely simplifying tree structures, such as depth limitations, inevitably leads to performance degradation. The second challenge concerns defining developer comprehensibility of algorithms. From a developer's perspective, algorithm comprehensibility is inherently subjective and cannot be easily optimized as an objective function~\cite{martens2011comprehensibilit}. Simple metrics such as tree depths and leaf node counts may inadequately capture all factors affecting developer understanding~\cite{huysmans2011comprehensibility}.

Therefore, we present \texttt{ComTree}, the first framework that incorporates comprehensibility into the generation process of bitrate adaptation algorithms.  \texttt{ComTree} aims to enhance algorithmic comprehensibility while maintaining performance through a two-phase approach: first constructing a comprehensive set of heterogeneous decision trees with comparable performance (known as the Rashomon set~\cite{fisher2019Rashomon2,semenova2022Rashomon1}), then conducting comprehensibility assessment using Large Language Models (LLMs)~\cite{zhao2023llmsurvey}.

To construct the Rashomon set, we first compress the ABR state space through feature elimination based on black-box model prior knowledge~\cite{mctavish2022gosdt-guess}, significantly reducing the time complexity of optimal decision tree generation. Using the compressed state space, we generate training datasets that align with actual playback probabilities through a teacher-student learning framework similar to previous work~\cite{bastani2018student-teacher,meng2019pitree}. Based on this dataset, we employ a thoroughly pruned dynamic programming algorithm~\cite{xin2022treefarms} to obtain a complete set of heterogeneous decision trees within a predetermined performance error range.
For the comprehensibility assessment phase, considering that LLMs have demonstrated human-like capabilities in various comprehension tasks through their learned world knowledge~\cite{achiam2023gpt,kaddour2023challenges}, we leverage LLMs as algorithm evaluators in place of developers. Inspired by the Human-Feedback methodology \cite{ouyang2022instructgpt}, we implement a pairwise comparison mechanism similar to human preference learning. 
To enhance evaluation reliability, we utilize an ensemble approach of LLMs~\cite{liang2023encouraging}, leveraging multiple LLMs to compare comprehensibility within the Rashomon set until consensus is reached. To eliminate the influence of similar decision trees, we employ a two-phase comparison method that achieves the theoretical minimum number of comparisons.

Our experimental results demonstrate that \texttt{ComTree} achieves superior comprehensibility while maintaining competitive performance compared to existing algorithms. In comparison with existing research~\cite{meng2019pitree}, \texttt{ComTree} achieves better performance using only approximately one-fifth of the number of nodes. It exhibits performance improvements ranging from 4\% to 98\% in average performance compared to classical baseline algorithms~\cite{bba,yin2015mpc,mao2017pensieve,xia2022genet,wu2024netllm,spiteri2020bola,meng2019pitree}, and ComTree's performance correspondingly improves as the teacher network's performance enhances~\cite{kan2022merina}. Furthermore, all decision trees in the Rashomon set generated by \texttt{ComTree} demonstrate strong competitive capability. Regarding comprehensibility evaluation, we found that ranking results from multiple LLMs exhibit a certain degree of consistency and reproducibility. Different runs within the same round show 60\% consistency, and 50\% of instances have final ranking differences within 10. Finally, we evaluated the performance of decision trees modified by LLMs in 5G scenarios, demonstrating the correlation between comprehensibility and the potential for further optimization by engineers.
\begin{figure*} 
  \centering
  \includegraphics[width=0.96\linewidth]{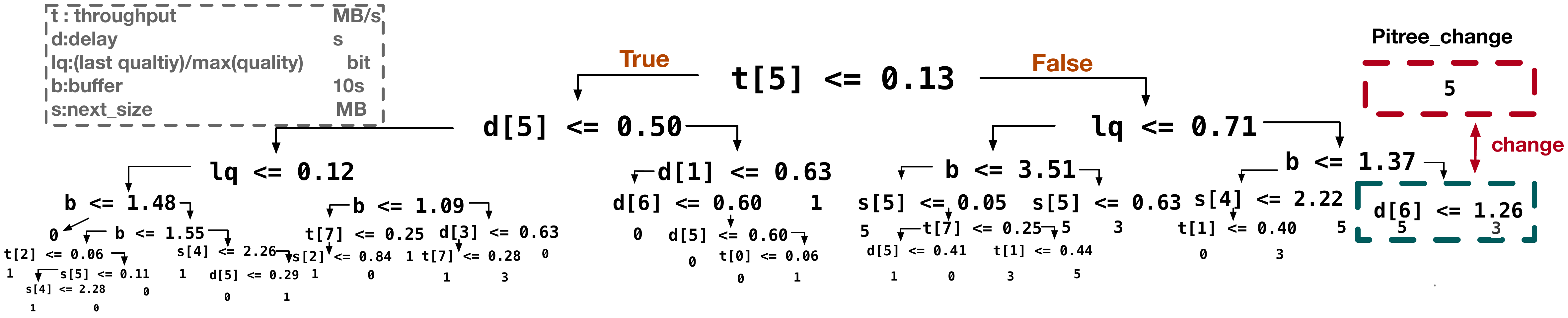}
    \vspace{-20pt}
  \caption{Decision Tree Generated by Pitree~(Some details omitted)
  }
  \vspace{-10pt}
  \label{fig:pitree}
\end{figure*}
\begin{table*}[h]
\centering
\caption{Comparison of Interpretability and Comprehensibility}
\vspace{-10pt}
\begin{tabular}{|p{2.2cm}|p{6.5cm}|p{3cm}|p{4.5cm}|}
\hline
\textbf{Feature} & \textbf{Focus} & \textbf{Hierarchy} & \textbf{Example} \\
\hline
Interpretability & Makes decision processes transparent. Addresses: \textbf{How did the model decide?} & Foundation for comprehensibility & Black-box model transformed into traceable decision tree \\
\hline
Comprehensibility & Focuses on understanding overall design logic. Addresses: \textbf{Why was the model designed this way?} & High-level goal related to developers' subjective understanding & Algorithms with high comprehensibility have greater potential for \textbf{improvement after adjustment} \\
\hline
\end{tabular}
\label{table:comparison}
\vspace{-10pt}
\end{table*}
In general, we summarize the contributions as follows:
\begin{itemize}[leftmargin=18pt]
\item We reveal the current limitations of adaptive video streaming in terms of comprehensibility and the importance of improving the comprehensibility of ABR through experimental validation.~(\S\ref{sec:mo})

\item We propose \texttt{ComTree}, the first work to introduce comprehensibility into adaptive video streaming. We utilize the Rashomon set and LLMs to find ABR algorithms that are more developer-friendly.~(\S\ref{sec:sys})

\item Through extensive experiments, we demonstrate that the algorithms generated by \texttt{ComTree} exhibit competitive performance in both performance and comprehensibility. Moreover, our framework shows potential for further optimization by engineers.~(\S\ref{sec:eva})
\end{itemize}

%% file: MOTIVATION.tex
\begin{figure}
    \centering
         \subfigure[FCC Network]{
        \includegraphics[width=0.46\linewidth]{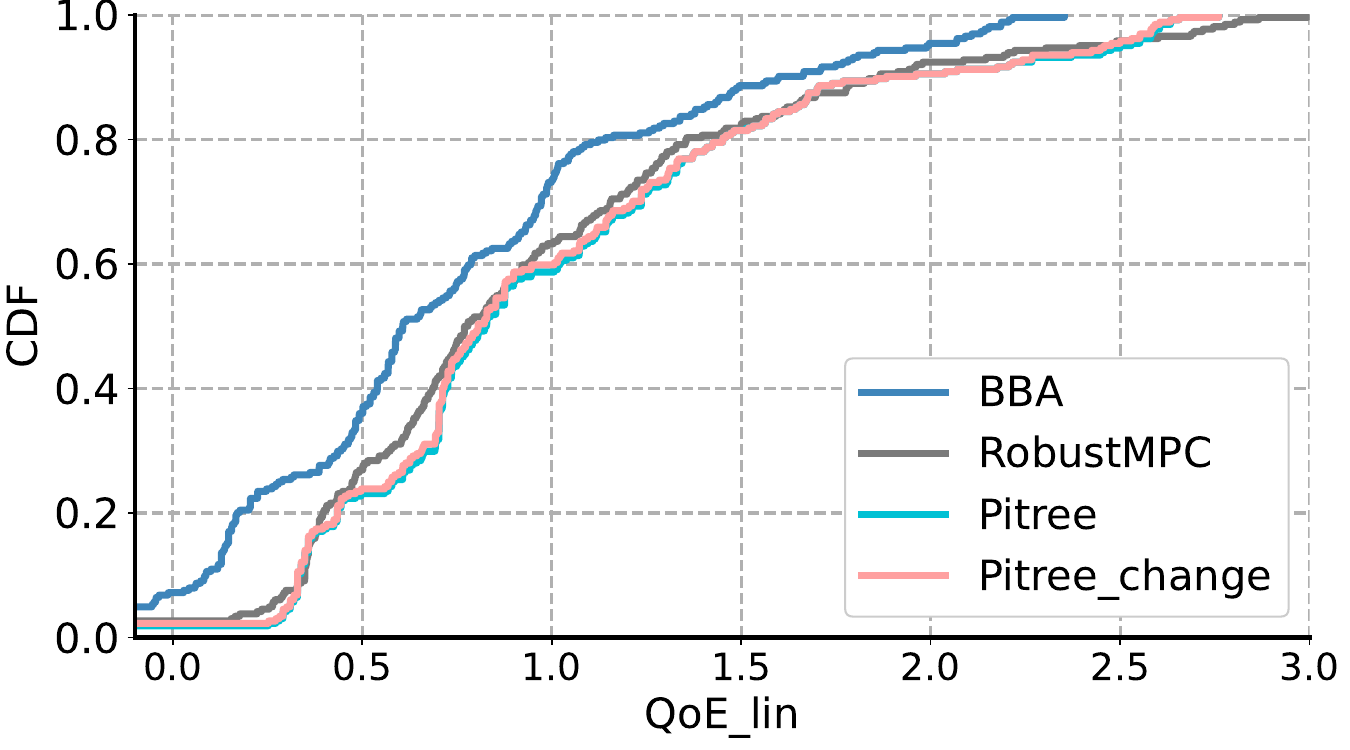}
        \label{fig:fcc}
        }
        \subfigure[5G Network]{
        \includegraphics[width=0.46\linewidth]{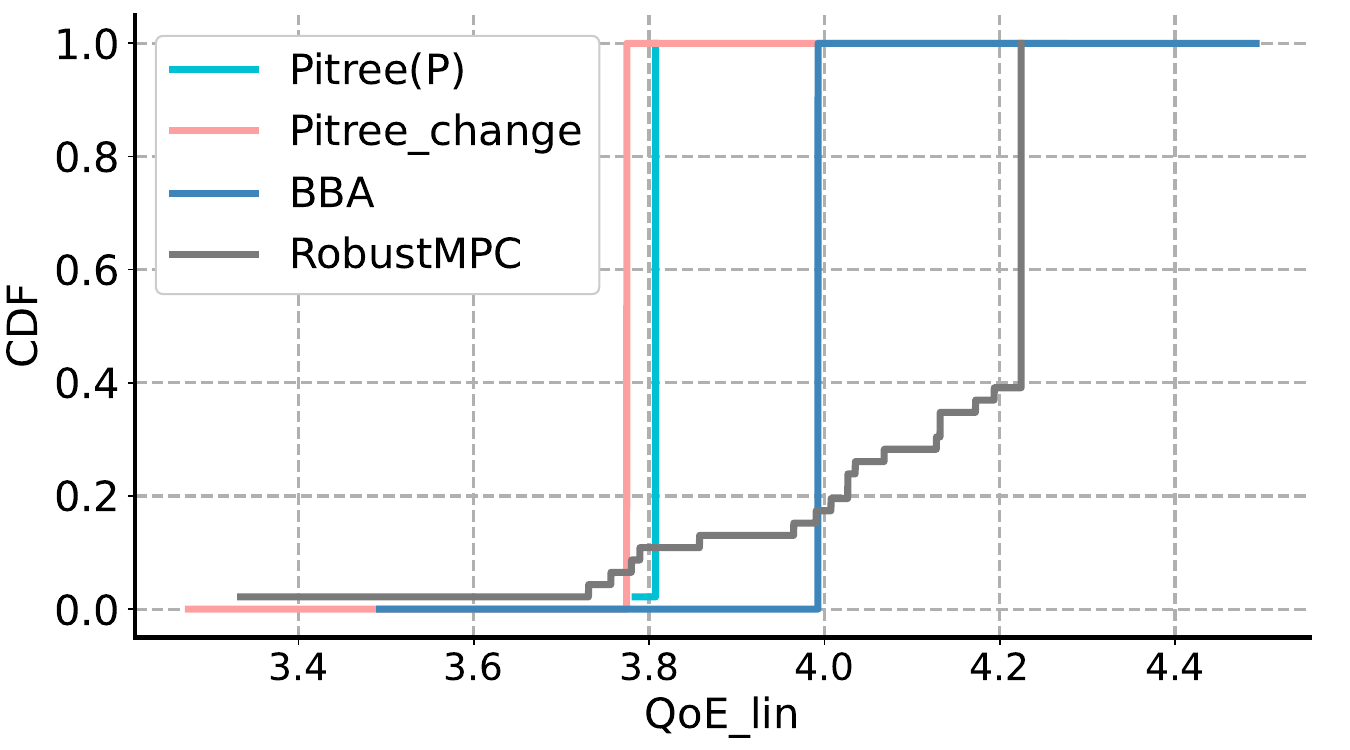}
        \label{fig:5g}
        }
         \vspace{-15pt}
         \caption{The CDF of $QoE_{lin}$ in FCC and 5G}
         \vspace{-15pt}
         \label{fig:The CDF of qoe}
\end{figure}

\begin{table}
\centering
\caption{Results of $QoE_{lin}$}
\vspace{-10pt}
\setlength{\tabcolsep}{1mm}
\begin{tabular}{ccccc}
\hline
&BBA&RobustMPC&Pitree&Pitree\_change\\
\hline
FCC&0.61 ($\pm$ 1.28)&0.83($\pm$ 1.30)&\textbf{ 0.88 ($\pm$ 1.30)}&0.87 ($\pm$ 1.30)\\
5G& 3.99 ($\pm$ 2e-15)&\textbf{4.12 ($\pm$ 0.19)}&3.81 ($\pm$ 3e-3)&3.77 ($\pm$ 8e-16)\\
\hline
\end{tabular}
\vspace{-15pt}
\label{table:moti}
\end{table}
\section{Background and Related Work}\label{sec:mo}
\subsection{Limitations: Inadequate Comprehensibility in Current Interpretable ABR Methods}
Comprehensibility, as a subjective metric, represents the degree to which users can understand an algorithm~\cite{freitas2014comprehensible,martens2011comprehensibilit,huysmans2011comprehensibility,freitas2006we,pazzani2000knowledge}. This characteristic plays a vital role in various domains such as finance~\cite{martens2007comprehensible} and medicine~\cite{huysmans2011comprehensibility}. We summarize the differences between interpretability and comprehensibility in Table~\ref{table:comparison}.
Previous ABR works~\cite{meng2019pitree,gruner2020reconstructing} transform black-box models into rule-based white-box models, but for developers, the current level of interpretability may be difficult to comprehend. Following the default parameters of Pitree~\cite{meng2019pitree} (FCC dataset~\cite{fcc}, 100 nodes, 500 iterations), we generate a decision tree, which is illustrated in Figure~\ref{fig:pitree}. The above tree has a total of 31 leaves, with a maximum depth of 8, and utilizes 14 distinct features. Compared to other heuristic algorithms~\cite{bba,yin2015mpc,spiteri2020bola}, it is challenging for developers to understand the design principles behind this tree, which poses challenges for further optimization.

We conduct tests on two datasets with different bandwidths: the low-bandwidth FCC dataset~\cite{fcc}~(average 1.31 Mbps, std 1.00) and the high-bandwidth 5G dataset~\cite{narayanan2020lumos5g} (average 347.46 Mbps, std 378.16). We compare Pitree, BBA, and RobustMPC~\cite{yin2015mpc}, and calculate the $QoE_{lin}$ as Table~\ref{table:moti}, showing the  Cumulative Distribution Function~(CDF) results in Figure~\ref{fig:The CDF of qoe}. Our experiments reveal that Pitree achieves optimal QoE in low-bandwidth scenarios. However, in high-bandwidth environments, its performance falls behind even basic algorithms like BBA. We observe that Pitree's suboptimal performance in high-bandwidth conditions can be attributed to its conservative approach, which results in delayed transitions to higher bitrates. Therefore, we attempt to modify Pitree as Figure~\ref{fig:pitree}. We remove the rightmost node (\texttt{d[6]<=1.26}), and when \texttt{b>1.37}, we output the highest bitrate for all cases. We name this modified version Pitree\_change. We expect that with our modifications, Pitree's performance will become more aggressive and achieve better results in high-bandwidth scenarios. However, the performance of Pitree\_change declines in both low-bandwidth and high-bandwidth datasets. This finding highlights that the current decision tree's complexity poses challenges for developers in terms of comprehension and maintenance. When confronted with performance degradation scenarios, the intricate structure of the decision tree makes it difficult for developers to implement timely and effective adjustments.

\subsection{Opportunity: From Large-scale Subjective Experiments to LLMs with Human-like Capabilities}
To assess the subjective metric of ABR comprehensibility, conducting large-scale user studies would be the most direct and reliable approach. However, this method faces significant challenges: it not only requires substantial time and cost investment but also demands participants with domain knowledge in ABR algorithms. Nevertheless, the recent rapid evolution of LLMs~\cite{nakano2021webgpt,achiam2023gpt,zhao2023llmsurvey} presents a promising opportunity to address these challenges. LLMs have demonstrated remarkable performance beyond traditional natural language processing tasks~\cite{raffel2020t5,devlin2019bert}, exhibiting strong human-like capabilities in accurately and efficiently simulating user behavior through their comprehensive general knowledge~\cite{kaplan2020scaling,brown2020gpt3}. For instance, several studies~\cite{park2023generative,aher2023using,hussain2024tutorial} have investigated human behavior simulation and subjective perception assessment powered by LLMs, generating credible individual and emergent social behaviors during evaluation. In the field of recommender systems, researchers have effectively utilized LLMs to simulate various user behaviors~\cite{ren2024bases,zhao2023recommender,huang2023recommender}, including browsing, searching, and viewing activities. Evaluation results consistently demonstrate that these simulated behaviors closely align with real human behavioral patterns.

%% file: system.tex
\section{ComTree Design}\label{sec:sys}

\begin{figure}
    \centering
        \includegraphics[width=0.95\linewidth]{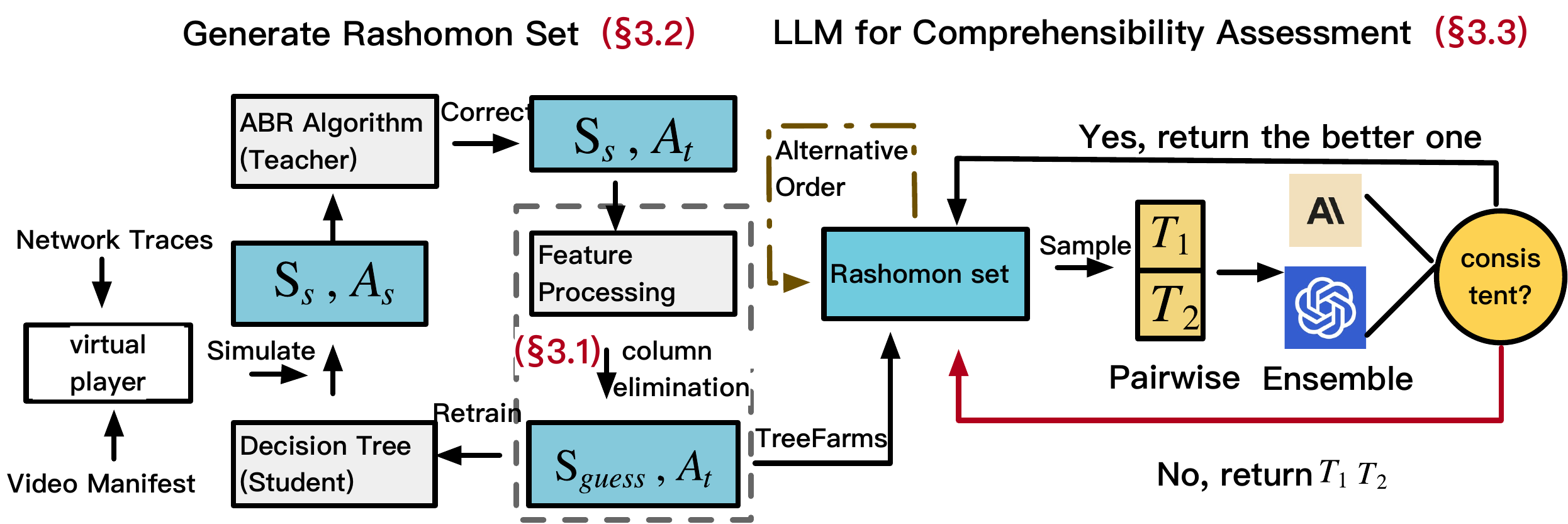} 
        \vspace{-5pt}
         \caption{System Overview}
         \vspace{-15pt}
         \label{System Overview}
\end{figure}

Our system, as illustrated in Figure~\ref{System Overview}, consists of two primary modules: the Rashomon Set Construction Module and the LLM Assessment Module. In the Rashomon Set Construction Module, we first introduce feature processing to reduce ABR state complexity~(\S\ref{sec:fp}), then employ a teacher-student learning framework to construct a dataset aligning with actual replay state probabilities and generate a Rashomon set meeting accuracy requirements~(\S\ref{sec:rset}). The LLM Assessment Module~(\S\ref{sec:llma}) defines comparison rules and interaction logic to select ABR with enhanced comprehensibility.

\subsection{Feature Processing}\label{sec:fp}
To construct the Rashomon set, we need to find the optimal tree via dynamic programming and then generate all trees meeting accuracy requirements~\cite{xin2022treefarms}. However, given the vast feature space in ABR, the dynamic programming method becomes computationally intractable as its complexity grows at a factorial level with the number of features~\cite{cormen2022introduction}. For example, in one training sample of Pitree, the training sample size is 200,000, and each state consists of 2 discrete features (last quality and chunk\_til\_video\_end) and 23 continuous features. After binary encoding, the size of the dataset becomes 200,000 x 704,661, which is difficult for a dynamic programming algorithm to handle.
    


    
To address this, we obtain knowledge gleaned from black-box models~\cite{mctavish2022gosdt-guess}. We use the guessing thresholds via column elimination to reduce the number of feature encodings. Specifically, we employ a parallelized XGBoost~\cite{chen2016xgboost} as the reference black-box algorithm. We first generate an ensemble tree using the original dataset. Then, to accelerate the iteration, we filter out features with importance less than $\delta$ ( We experimentally set $\delta$  as 1e-4). Next, we remove the feature with the lowest importance and regenerate the ensemble tree. If the accuracy of the newly generated ensemble tree is not lower than that of the previous trees, it indicates that the feature can be eliminated without affecting accuracy. We iterate this step until the accuracy of the generated ensemble tree no longer meets the requirements. Through iteration, the number of encoded features is reduced from 704,661 to 40-60, greatly reducing the time required to find the optimal tree. The details can be found in Appendix~\ref{app:alg}.


\subsection{Generate the Rashomon Set}\label{sec:rset}
 Feature processing significantly reduces ABR state complexity, creating space for constructing the Rashomon set. We first employ a teacher-student learning framework to obtain a dataset that reflects realistic playback probabilities~\cite{bastani2018student-teacher,meng2019pitree}. Specifically, we use the data generated by rolling out the initial black-box teacher network to train the initial version of the student decision tree. Subsequently, we deploy the student decision tree in a simulated video playback environment~\cite{yin2015mpc}. The decision tree interacts with the virtual player under specified network conditions, generating student network execution states $s_s$ and actions $a_s$. After the interaction is completed, we provide $s_s$ to the teacher network, which generates the teacher action $a_t$. The pairs of student states and teacher actions ($s_s$, $a_t$) are then saved in the dataset, and in subsequent training, the previous actions are corrected. This approach avoids cascading errors and yields a set of states and teacher actions that conform to the real playback probability.

After obtaining a dataset with realistic playback probabilities through the teacher-student learning framework, we establish conditions for generating a complete set of heterogeneous equivalent decision trees. To generate the Rashomon set, we employ TreeFarms~\cite{xin2022treefarms}, a dynamic programming algorithm based on the Generalized and scalable optimal sparse decision trees~(GOSDT)~\cite{lin2020gosdt} with comprehensive compression and pruning mechanisms. The optimization objective of GOSDT combines misclassification loss with a sparsity penalty on the number of leaves, expressed as:
\begin{equation}
obj  = loss_{mis} + \lambda H_{t}
\label{opt_gosdt}
\vspace{-5pt}
\end{equation} where $H_{t}$ is
the number of leaves in tree $t$ and $\lambda$ is a regularization parameter. For the objective of the optimal tree $obj_{opt}$ found by GOSDT, TreeFarms calculates the bound threshold $\theta_{\epsilon}$ of the Rashomon set following:
\begin{equation}
\theta_{\epsilon} = obj_{opt} * (1 + \epsilon )
\label{opt_tree}
\vspace{-5pt}
\end{equation}
 Specifically, for each subproblem in the dynamic program, TreeFarms keeps track of upper and lower bounds on its objective and removes only those subproblems whose objective lower bound is greater than $\theta_{\epsilon}$. It stores these subproblems and their bounds in a dependency graph, which expresses the relationships between subproblems. Finally, it returns all models within $\theta_{\epsilon}$. The overall framework of the Rashomon set construction algorithm is presented in Appendix~\ref{app:alg}.

\begin{algorithm}
    \caption{LLM-based Comprehensibility Assessment}
    \label{alg:llm}
    \KwIn{Rashomon set $R_{set}$, LLMs $f_{GPT}$,$f_{Claude}$}
    \KwOut{Most comprehensible decision tree $t_{opt}$}
    
    \While{$|R_{set}| > 1$}
    {
         $R'_{set} = shuffle(R_{set})$ \\
        {\color{blue}/* Pairwise comparison phase */} \\
        \For{$i \in [0, \lfloor(|R'_{set}|-1)/2\rfloor]$}{
            $T_i = R'_{set}[2i]$, $T_j = R'_{set}[2i+1]$ \\
            {\color{teal}// Ensemble evaluation using both LLMs} \\
            $r_1 = f_{GPT}(T_i, T_j)$, $r_2 = f_{Claude}(T_i, T_j)$ \\
            \If{$r_1 == r_2$}{
                Remove less comprehensible tree from $R'_{set}$ \\
            }
        }
        
        {\color{blue}/* Alternative comparison order */} \\
        \If{$R_{set} == R'_{set}$}{
            \For{$i \in [1,\lfloor(|R'_{set}|/2\rfloor]$}{
                $T_i = R'_{set}[i*2]$, $T_j = R'_{set}[i*2-1]$ \\
                $r_1 = f_{GPT}(T_i, T_j)$, $r_2 = f_{Claude}(T_i, T_j)$ \\
                \If{$r_1 == r_2$}{
                    Remove less comprehensible tree from $R'_{set}$ \\
                }
            }
            {\color{teal}// Local optimum detection} \\
            \If{$R_{set} == R'_{set}$}{
                \Return $R_{set}$ \\
            }
        }
        $R_{set} = R'_{set}$ \\
    }
    $t_{opt} = R_{set}$ \\
    \Return $t_{opt}$
\end{algorithm}
\subsection{Comprehensibility Assessment Utilizing LLMs}\label{sec:llma}
We propose a framework for evaluating the Rashomon set comprehensibility based on LLMs. The framework incorporates three core designs: a pairwise comparison mechanism that leverages LLMs' strength in relative judgments; a model ensemble strategy that enhances evaluation reliability through multi-model consensus; and a two-stage comparison mechanism that addresses model inconsistencies and similar decision trees with the theoretical minimum number of comparisons.

\textbf{Pairwise Comparison} Our framework leverages LLMs' understanding and reasoning capabilities to automate the evaluation of decision tree comprehensibility. We adopt a pairwise comparison mechanism, grounded in LLMs' training process: through human preference alignment~\cite{ouyang2022instructgpt}, LLMs learn to understand human subjective judgments through relative ranking of alternatives. Therefore, employing pairwise comparisons rather than direct scoring aligns better with their pre-training paradigm and better utilizes their capability to understand human subjective judgments.

\textbf{Model Ensemble} In implementation, we employ a model ensemble strategy~\cite{liang2023encouraging} incorporating multiple LLMs (e.g., GPT-4o~\cite{GPT4o} and Claude-3.7-Sonnet~\cite{Claude}). This multi-model ensemble approach not only mitigates individual model randomness and bias but also enhances evaluation credibility through consensus checks. Formally, for any two decision trees $T_i$ and $T_j$, we only eliminate the less comprehensible tree when multiple models reach consensus on comprehensibility judgment.

\textbf{Two-Phase Comparison}~In the evaluation process, when the comprehensibility of certain decision trees is very similar, forcing a distinction may lead to unstable results. To address this issue and determine whether the comprehensibility is consistent across the entire set or just within current groups, we designed a two-phase comparison mechanism based on the transitivity of relationships, achieving the theoretical minimum number of comparisons while ensuring evaluation quality.

Specifically, for a set of decision trees $R_{set}^{'}$, in the first phase, the algorithm performs pairwise comparisons after grouping: for $i \in \{0,1,2,...,\lfloor \frac{n-1}{2} \rfloor\}$,
the LLMs evaluate the relative comprehensibility of decision tree pairs $(R_{set}^{'} (2i), R_{set}^{'} (2i+1))$. Based on the comparison results, trees with similar comprehensibility are returned to the set. 
If the first phase fails to establish any relative relationships, the algorithm proceeds to the second phase: for $i \in \{1,2,...,\lfloor \frac{n}{2} \rfloor\}$, the LLMs further evaluates decision tree pairs $(R_{set}^{'} (2i), R_{set}^{'} (2i-1))$. If certain decision trees remain indistinguishable after these two phases, the transitivity of comparison relationships implies that these trees have sufficiently similar comprehensibility and should be classified into the same equivalence class. Furthermore, the two-phase comparison precisely provides n-1 independent comparisons forming a minimum spanning tree, ensuring both connectivity and minimality~\cite{cormen2022introduction} across all decision trees. The complete algorithmic workflow of the evaluation framework is detailed in Algorithm~\ref{alg:llm}.

    


    

%% file: evaluation.tex
\section{EVALUATION}~\label{sec:eva}
We introduce implementation details and experiment setup in \S\ref{exsetup} and evaluate \texttt{ComTree} by following aspects:
\begin{itemize}[leftmargin=8pt]
\item \textit{\textbf{Performance of the Optimal Tree}} We enhance our decision tree generation algorithm from a greedy approach to a dynamic programming algorithm with optimality guarantees, while \texttt{ComTree} maintains only one-fifth of the previous nodes~\cite{meng2019pitree}. Through comprehensive evaluations across diverse network trace-driven simulations and real-world environments, as demonstrated in \S\ref{sec:eva_opt}, our optimal tree demonstrates competitive performance while maintaining a reduced tree size.

\item \textit{\textbf{Performance of the Rashomon Set}} We analyze the characteristics of decision trees within the Rashomon set and demonstrate the competitive performance distribution of the Rashomon set under different network traces as shown in \S\ref{sec:eva_rset}.

\item \textit{\textbf{Comprehensibility}} We illustrate the LLM selection process across different rounds and runs, analyzing the selection criteria as presented in \S\ref{sec:eva_inter}. Finally, through experiments in novel network environments substantially different from the training set, we demonstrate that the most comprehensible ABR algorithm selected by \texttt{ComTree} shows greater potential for further improvements, as detailed in \S\ref{sec:eva_potential}.

\end{itemize}
We compare \texttt{ComTree} with Meta-RL algorithms~(Merina~\cite{kan2022merina}) and parameter adaptation algorithms(Oboe~\cite{akhtar2018oboe}) in Appendix~\ref{app:com} to further discuss the positioning of \texttt{ComTree}, and discuss the limitations and future work directions of \texttt{ComTree} in Appendix~\ref{app:dis}.
\begin{table}
\centering
\caption{Parameters in $QoE$}
\vspace{-10pt}
\begin{tabular}{lll}
\hline
$QoE_{lin}$ & $q(R)=R$ & $\mu=4.3$ \\
\hline
$QoE_{hd}$ & $q(R)=\{0.3:1, 0.75:2, 1.2:3,$ & $\mu=8$ \\
&\hspace{1cm}$1.85:12, 2.85:15, 4.3:20\}$ & \\
\hline
\end{tabular}

\vspace{-5pt}
\label{table:qoe}
\end{table}



\begin{figure*}
    \centering
      \subfigure[Norway]{
        \includegraphics[width=0.23\linewidth]{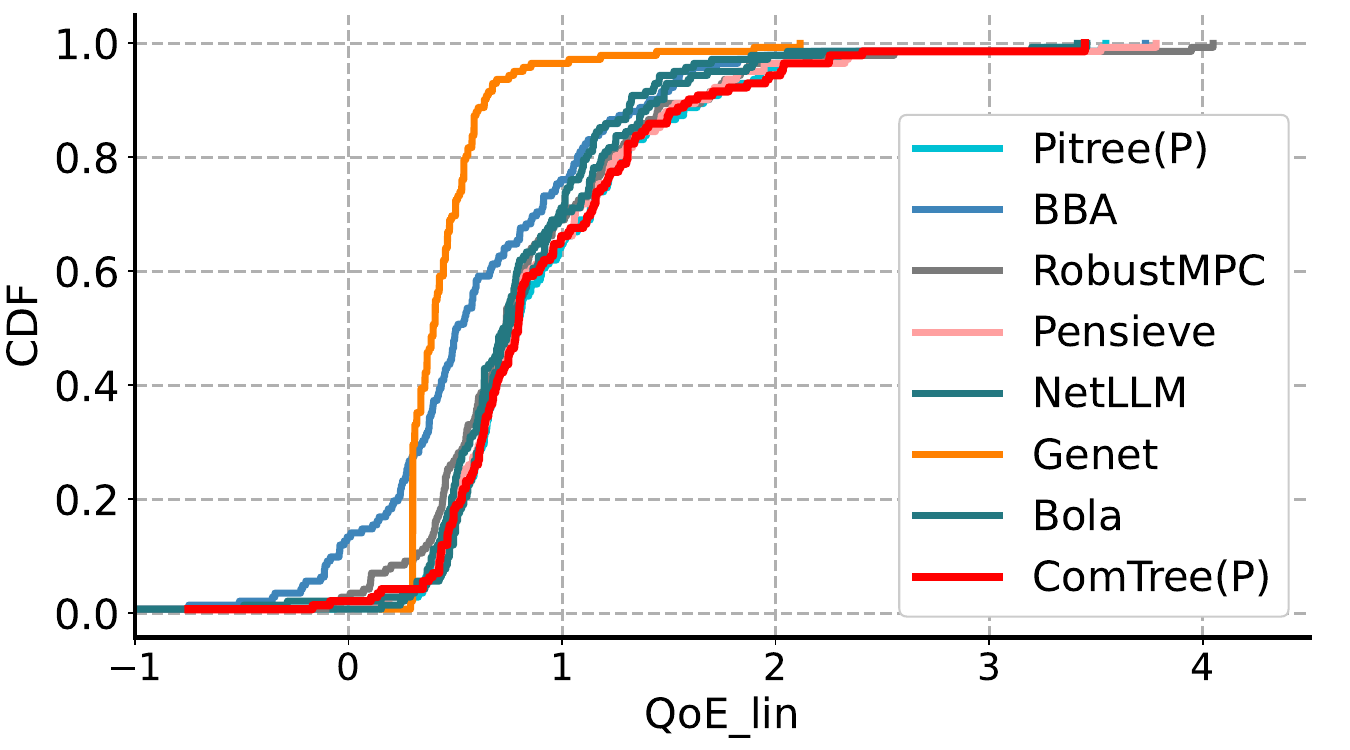}
        \label{fig:exp1nor}
        }
         \subfigure[Oboe]{
        \includegraphics[width=0.23\linewidth]{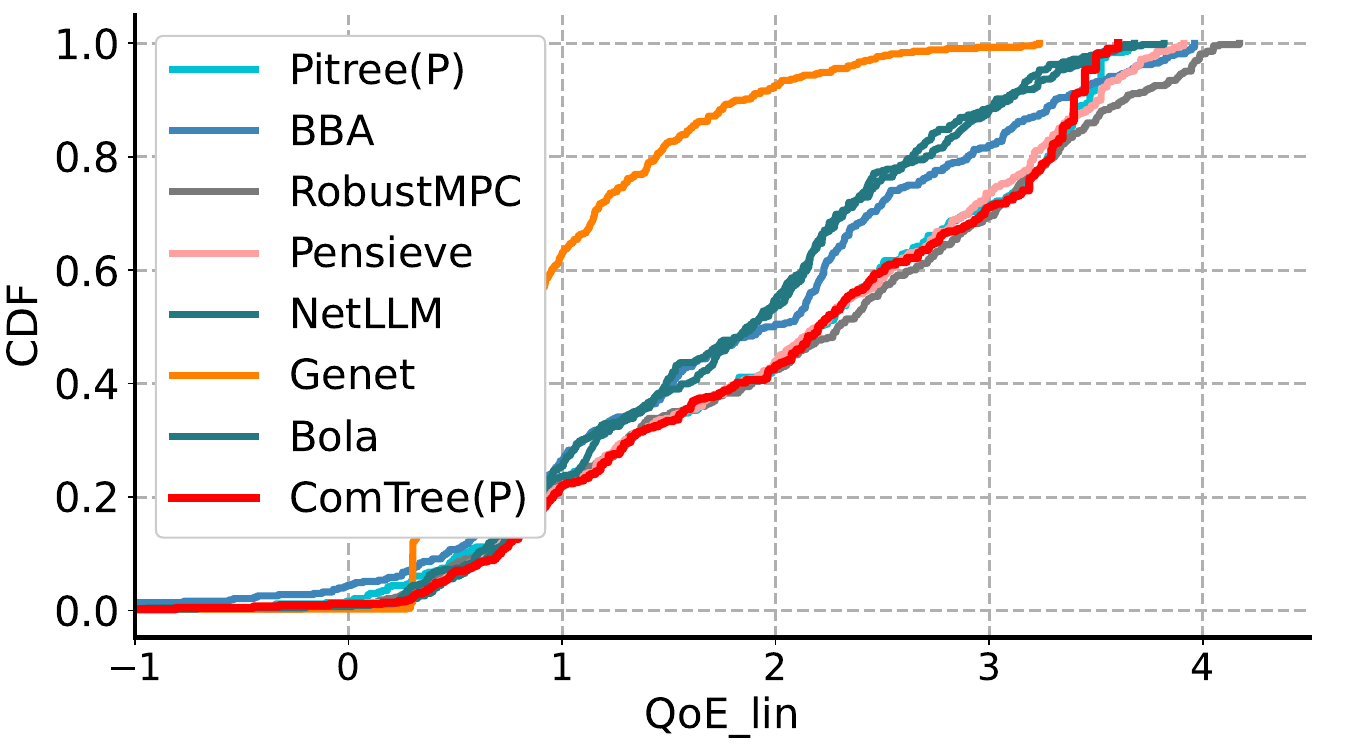}
        \label{fig:exp1oboe}
        }
      \subfigure[Puffer-Oct.17-21]{
        \includegraphics[width=0.23\linewidth]{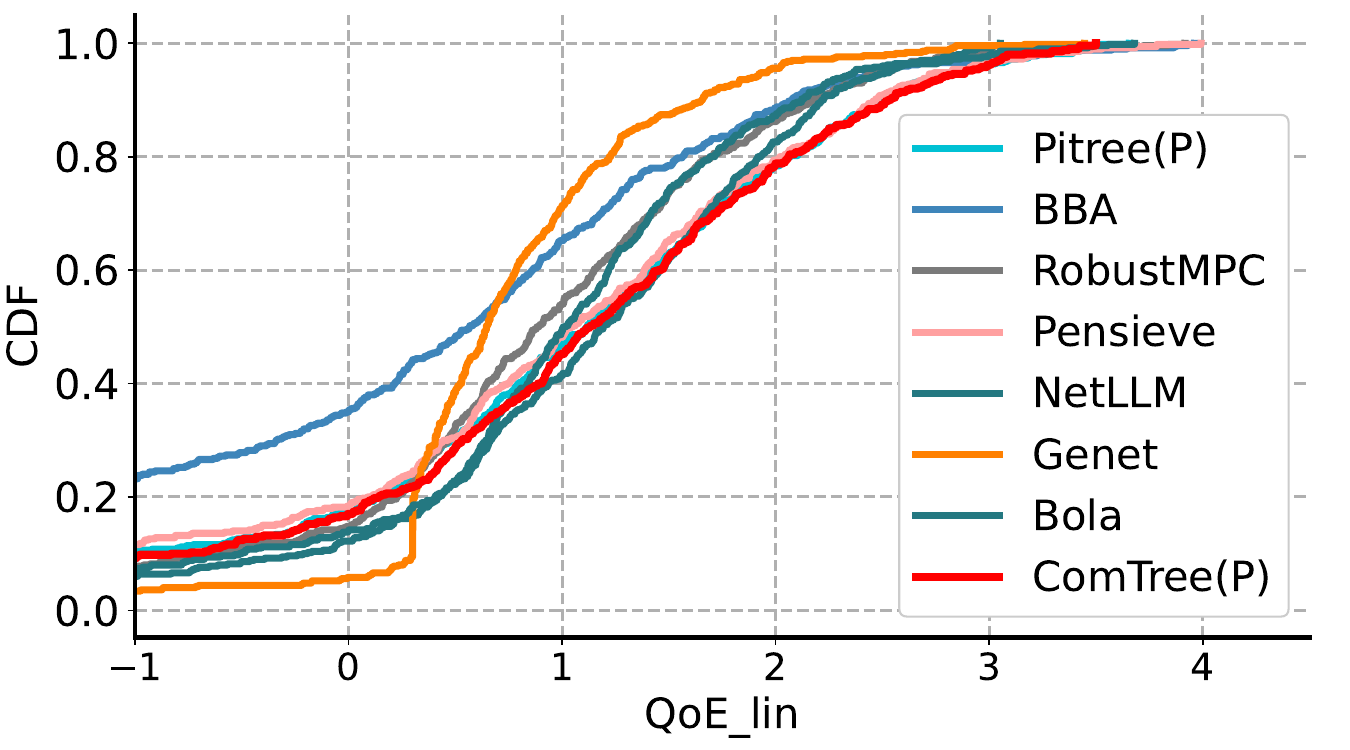}
        \label{fig:exp1puffer1}
        }
      \subfigure[Puffer-Feb.18-22]{
        \includegraphics[width=0.23\linewidth]{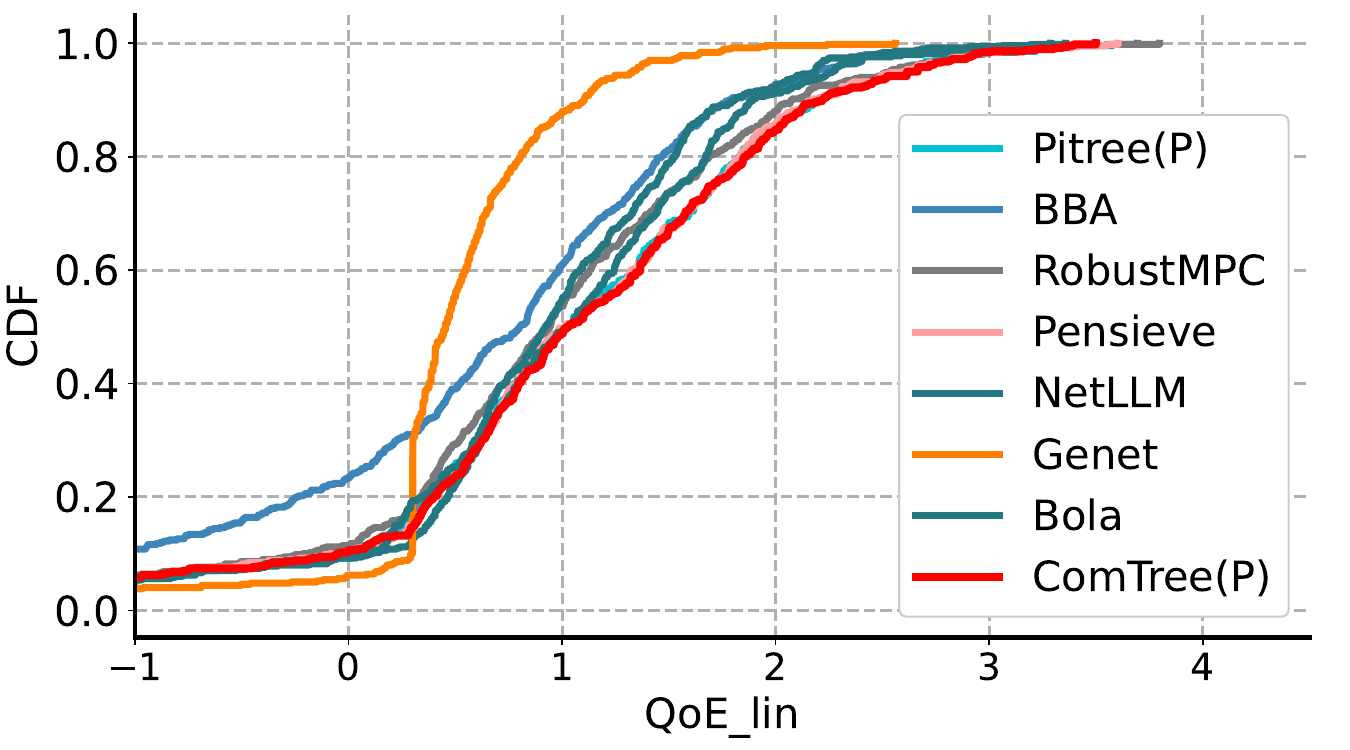}
        \label{fig:exp1puffer2}
        }
    
        \vspace{-15pt}
        \caption{CDF of $QoE_{lin}$ in different traces}
        \label{fig: sim}
        \vspace{-5pt}
\end{figure*}

\begin{table}
\centering
\caption{Details of \texttt{ComTree} in Evaluation}
\vspace{-10pt}
\begin{tabular}{l|l}
\hline
\texttt{ComTree}(P)& The optimal tree generated by teacher Pensieve\\
\hline
\texttt{ComTree}\_C&The most comprehensible  tree in Rashomon\\
\hline
'-L'  suffix& Adjusted by LLM

\end{tabular}
\vspace{-15pt}
\label{table:detail}
\end{table}

\begin{table*}
\centering
\caption{$QoE_{lin}$ in Different Traces}
\vspace{-10pt}
\setlength{\tabcolsep}{0.8mm}
\begin{tabular}{l|llllllll}
& \texttt{ComTree}(P) & Pitree(P) & RobustMPC & NetLLM & Bola & Pensieve & Genet & BBA \\ \hline
Norway & \textbf{0.93}$^\dagger$ ($\pm$ 0.58) & \textbf{0.94}$^*$ ($\pm$ 0.57) & 0.85 ($\pm$ 0.64) & 0.87 ($\pm$ 0.52) & 0.83 ($\pm$ 0.47) & \textbf{0.92}$^\ddagger$ ($\pm$ 0.56) & 0.46 ($\pm$ 0.25) & 0.64 ($\pm$ 0.65) \\
Oboe & \textbf{2.11}$^\ddagger$ ($\pm$ 1.06) & 2.08 ($\pm$ 1.10) & \textbf{2.16}$^*$ ($\pm$ 1.15) & 1.81 ($\pm$ 0.90) & 1.80 ($\pm$ 0.93) & \textbf{2.10}$^\dagger$ ($\pm$ 1.08) & 0.95 ($\pm$ 0.62) & 1.84 ($\pm$ 1.12) \\
Puffer-2110 & \textbf{0.76}$^\ddagger$ ($\pm$ 2.34) & 0.72 ($\pm$ 2.36) & 0.68 ($\pm$ 1.85) & \textbf{0.83}$^\dagger$ ($\pm$ 2.12) & \textbf{0.79}$^*$ ($\pm$ 1.81) & -0.13 ($\pm$ 12.67) & 0.63 ($\pm$ 1.30) & -0.18 ($\pm$ 2.92) \\
Puffer-2202 & \textbf{0.78}$^*$ ($\pm$ 2.98) & \textbf{0.73}$^\dagger$ ($\pm$ 3.05) & \textbf{0.67}$^\ddagger$ ($\pm$ 2.92) & 0.66 ($\pm$ 3.22) & 0.60 ($\pm$ 3.01) & 0.29 ($\pm$ 10.47) & 0.25 ($\pm$ 2.85) & 0.25 ($\pm$ 3.18) \\
\hline
Average & \textbf{1.15}$^*$~\textbf{(1st)} & \textbf{1.11}$^\dagger$~\textbf{(2nd)} & \textbf{1.10}$^\ddagger$~\textbf{(3rd)} & 1.04~\textbf{(4th)} & 1.00~\textbf{(5th)} & 0.71~\textbf{(6th)} & 0.58~\textbf{(7th)} & 0.58~\textbf{(8th)} \\
\end{tabular}
\vspace{-15pt}
\label{table:sim}
\end{table*}

\subsection{Experiment Setup and Implementation Details }~\label{exsetup}

\subsubsection{Video Sample} We use the same "EnvivoDash3" video from the MPEG-DASH reference videos as previous work~\cite{mao2017pensieve,meng2019pitree}, with a length of 193 seconds. The video is divided into 4-second segments, with bitrates of {300, 750, 1200, 1850, 2850, 4300} kbps.
\subsubsection{QoE Metrics} The QoE metric can be expressed by video quality, rebuff penalty and smooth penalty as:
\begin{equation}
QoE = \sum_{n}q(R_n) - \mu \sum_{n}T_n - \sum_{n}|q(R_{n+1} )-q(R_{n})|
\label{qoe}
\vspace{-5pt}
\end{equation}
Here, $q(R_n)$ represents the video quality when the $n$-th video segment selects bitrate level $R$, and $T_n$ represents the rebuffering duration for the $n$-th segment.
We primarily use $QoE_{lin}$ as the evaluation metric, as it is one of the most commonly used evaluation indicators~\cite{yin2015mpc,mao2017pensieve}. To verify the algorithm's performance under different QoE metrics, we also use $QoE_{hd}$ as another evaluation metric in Figure~\ref{fig:QoE ratio}. Specifically, the details of each QoE metric as Table~\ref{table:qoe}. To further illustrate the performance difference between \texttt{ComTree} and other baseline methods, we employ the QoE Improvement Ratio, denoted as $QoE_{impro}$:
\begin{equation}
QoE_{impro} = (QoE - QoE_{baseline}) / |QoE_{baseline}|
\label{qoe_ratio}
\end{equation}

\subsubsection{Network Trace Datasets} We leverage network traces for training and testing in the simulation environment. Specifically, we use the FCC~\cite{fcc} dataset for training. Then, in \S\ref{sec:eva_opt} and \S\ref{sec:eva_rset}, we employ the Norway~\cite{riiser2013norway}, Oboe~\cite{akhtar2018oboe}, and two datasets from the Puffer~\cite{yan2020fugu} platform for testing. In \S\ref{sec:eva_inter}, we utilize the 5G~\cite{narayanan2020lumos5g} dataset, which differs significantly from the training trace FCC, to demonstrate the comprehensibility potential of \texttt{ComTree}.

\subsubsection{ABR Baselines} 
We select representative algorithms across various design paradigms as baselines. For heuristic approaches, we include buffer-based algorithms \textbf{BBA}~\cite{bba} and \textbf{Bola}~\cite{spiteri2020bola}, along with the model-based algorithm \textbf{RobustMPC}~\cite{yin2015mpc}. Among learning-based methods, we choose \textbf{Pensieve}~\cite{mao2017pensieve} based on deep reinforcement learning, \textbf{Genet}~\cite{xia2022genet} utilizing curriculum learning, and \textbf{NetLLM}~\cite{wu2024netllm} leveraging large language models. From previous interpretability-focused work, we include \textbf{Pitree}~\cite{meng2019pitree}, an algorithm that transforms black-box models into decision trees, and implement \textbf{Pitree(P)} using Pensieve as the teacher network.

\subsubsection{Implementation Details of \texttt{ComTree}} 
We configure the importance regularization parameter $\lambda$ as 0.0005, set the Rashomon set boundary $\epsilon$ to 0.05, and limit the maximum tree depth $d$ to 6. All experiments are conducted on Ubuntu 18 with dual AMD EPYC 7742 processors. In Section~\ref{sec:eva_opt}, we generate the optimal decision tree using Pensieve as the teacher network, denoted as \textbf{\texttt{ComTree}(P)}. For Sections~\ref{sec:eva_rset} and~\ref{sec:eva_inter}, we select 64 distinct instances from the Rashomon set, sorted by objective values, to analyze the set's properties and evaluate the LLM assessment process and results. In Section~\ref{sec:eva_potential}, we apply LLM-based adjustments (denoted by the suffix -L) to the most comprehensible tree from the Rashomon set (\textbf{\texttt{ComTree}\_C}), the previously optimal tree (\texttt{ComTree}(P)), and Pitree (\textbf{Pitree(P)}). The differences are presented in Table~\ref{table:detail}.

\subsection{Performance of The Optimal Tree}~\label{sec:eva_opt}
\subsubsection{Trace-driven Simulation Experiment}
We run the simulated play on four datasets: Norway, Oboe, Puffer-Oct. 17-21 and Puffer-Feb. 18-22. Figure~\ref{fig: sim} illustrates the CDF curves of $QoE_{lin}$, while Table~\ref{table:sim} presents the mean and variance of $QoE_{lin}$ for each algorithm across different datasets. \texttt{ComTree}(P) consistently maintains top-three performance across multiple datasets, achieving optimal results in terms of average performance with improvements ranging from 4\% to 98\% compared to classical baseline algorithms.

\begin{figure}
    \centering
         \subfigure[\texttt{ComTree}(P) Compared with Pensieve]{
        \includegraphics[width=0.46\linewidth]{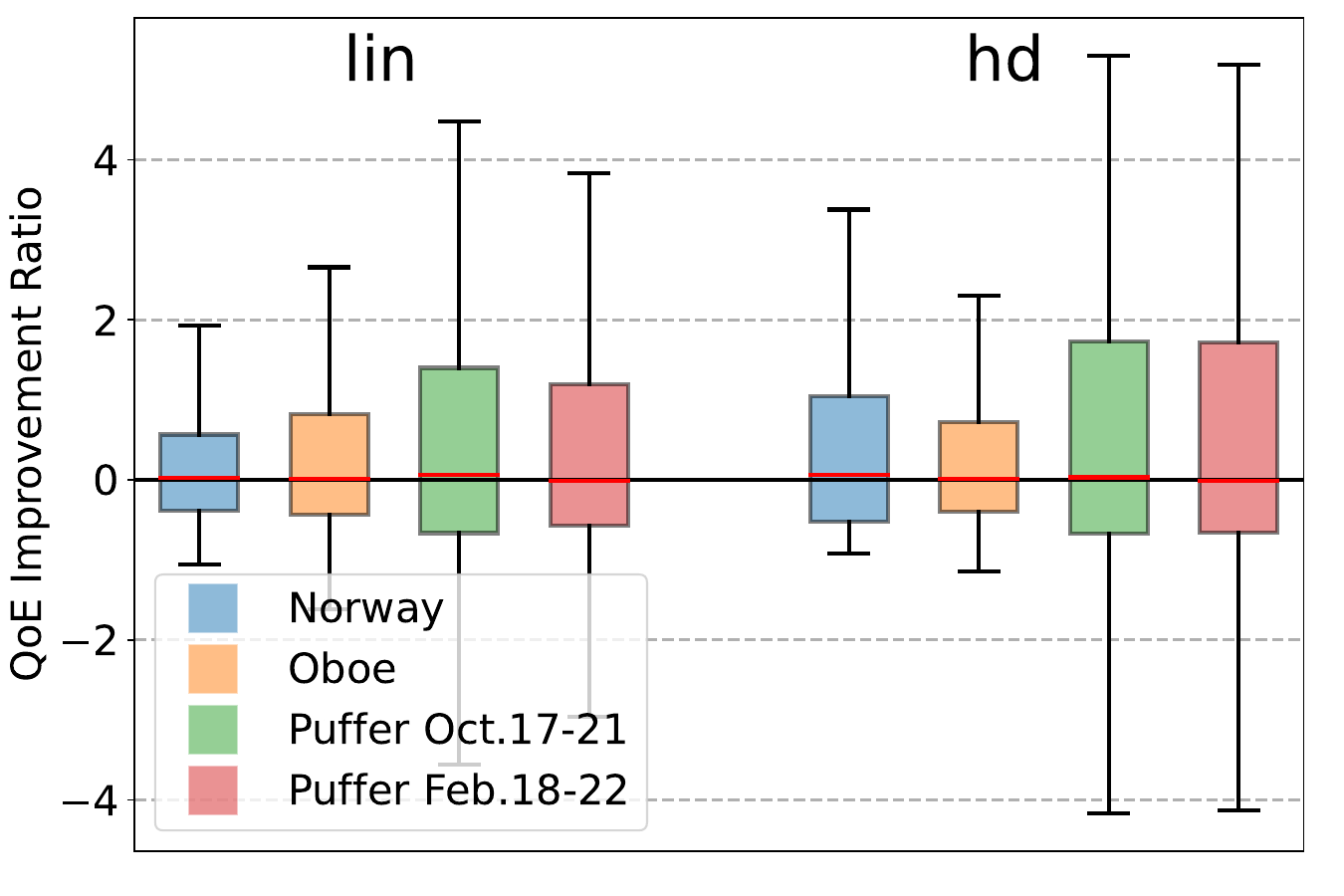}
        \label{fig:box/rl}
        }
        \subfigure[\texttt{ComTree}(P) Compared with Pitree(P)]{
        \includegraphics[width=0.46\linewidth]{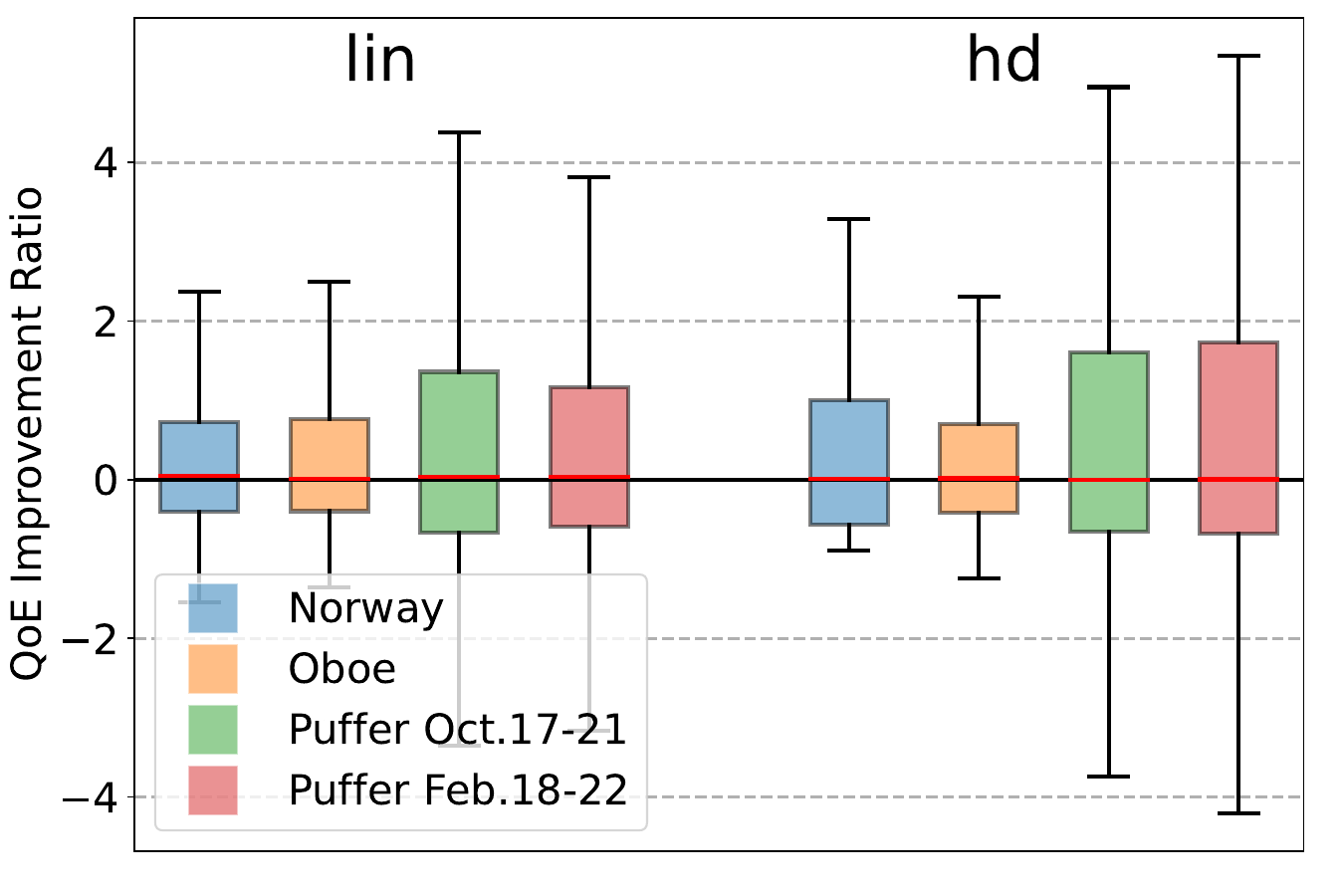}
        \label{fig:box/trl}
        }
         \vspace{-15pt}
         \caption{Improvement Ratio of \texttt{ComTree} on Different ABR Algorithms, Network Traces and QoE Metrics}
         \vspace{-15pt}
         \label{fig:QoE ratio}
\end{figure}

We analyze each trace performance using $QoE_{impro}$ combined with $QoE_{lin}$ and $QoE_{hd}$ metrics, as shown in Figure~\ref{fig:QoE ratio}. The comparison with the teacher algorithm (Figure~\ref{fig:box/rl}) and Pitree (Figure~\ref{fig:box/trl}) shows the upper box plot regions exceeding the lower regions, indicating that \texttt{ComTree} demonstrates advantages across various network tracking and QoE metrics. Compared to Pitree, \texttt{ComTree} employs dynamic programming with optimality guarantees rather than greedy algorithms. In contrast to the teacher algorithm Pensieve, \texttt{ComTree} achieves smoother decision boundaries by extracting key decision logic.

\begin{figure}
    \centering
           \includegraphics[width=0.96\linewidth]{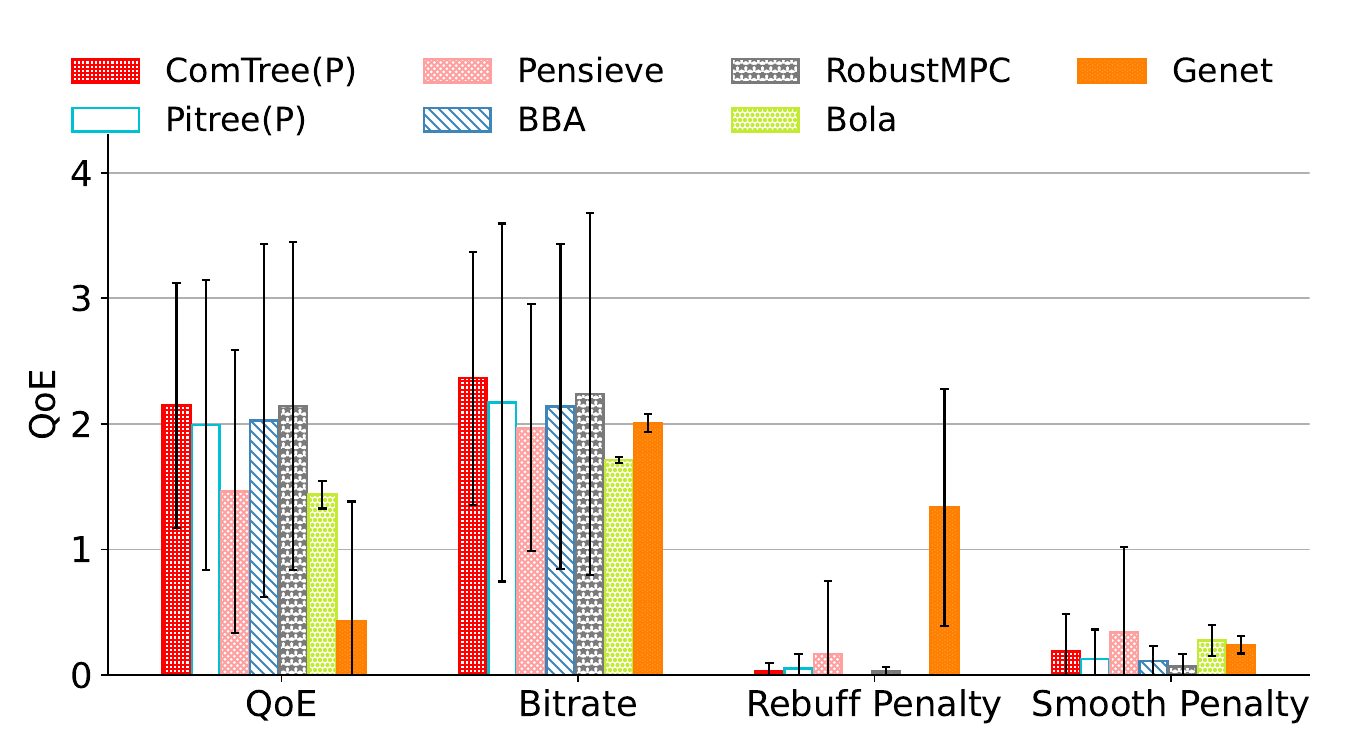}
         \vspace{-10pt}
         \caption{$QoE_{lin}$ and its Components in Real-World}
         \vspace{-15pt}
         \label{fig:real_world}
\end{figure}

\subsubsection{Real-world Experiments}
We integrate these algorithms into dash.js~\cite{dashjs} and conduct playback experiments using Selenium-automated browser testing.  The client are connected to a public WiFi network, with each algorithm randomly executed three times for sessions lasting 5 minutes each. Experimental results are illustrated in Figure~\ref{fig:real_world}. \texttt{ComTree}(P) demonstrates exceptional performance, achieving the highest scores in both the $QoE_{lin}$ metric and bitrate, while also outperforming the teacher algorithm in stall and smoothness metrics.

\subsection{Characteristics and Performance of the Rashomon Set}\label{sec:eva_rset}

\begin{figure}
    \centering
        \subfigure[ Number of Tree in Different Instance]{
        \includegraphics[width=0.46\linewidth]{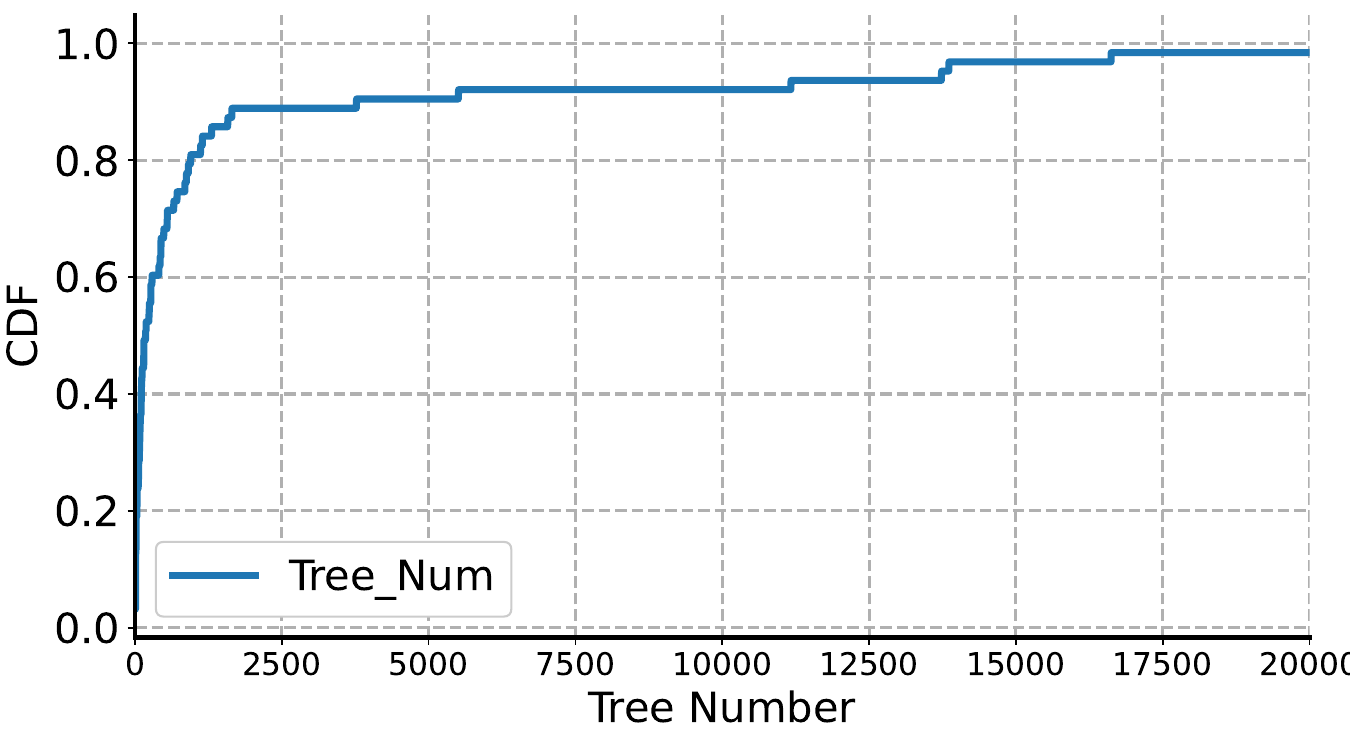}
        \label{fig:number}
        }
         \subfigure[ Feature Utilization Rate ]{
        \includegraphics[width=0.46\linewidth]{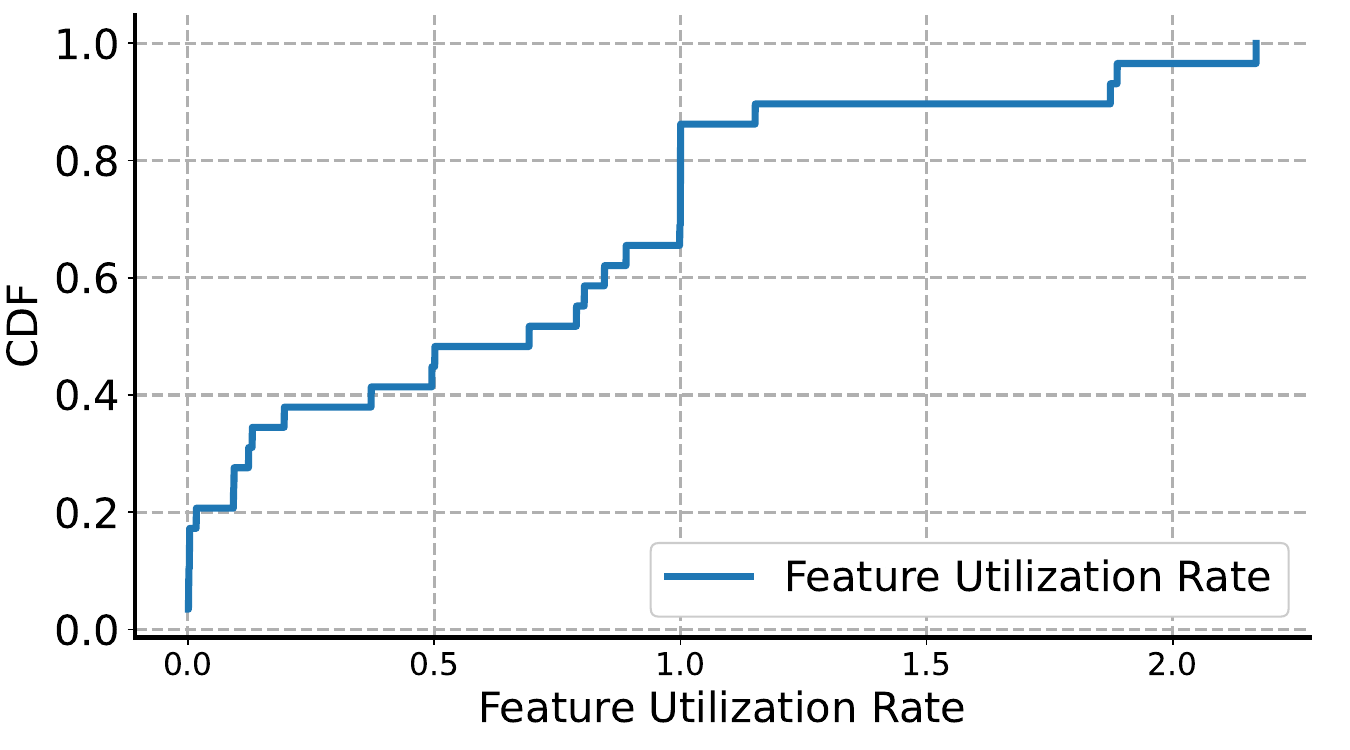}
        \label{fig:feature}
        }
         \subfigure[ Number of leaf Nodes ]{
        \includegraphics[width=0.46\linewidth]{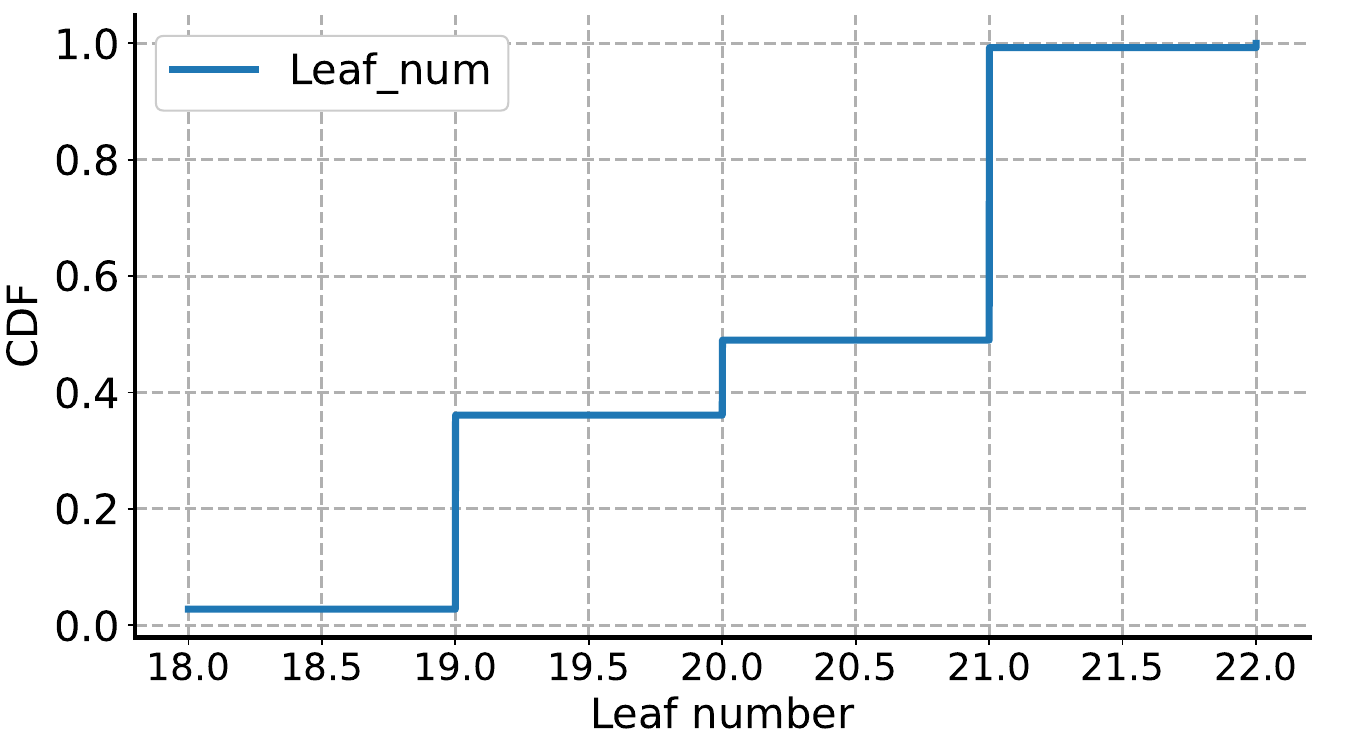}
        \label{fig:leaf}
        }
        \subfigure[Accuracy]{
        \includegraphics[width=0.46\linewidth]{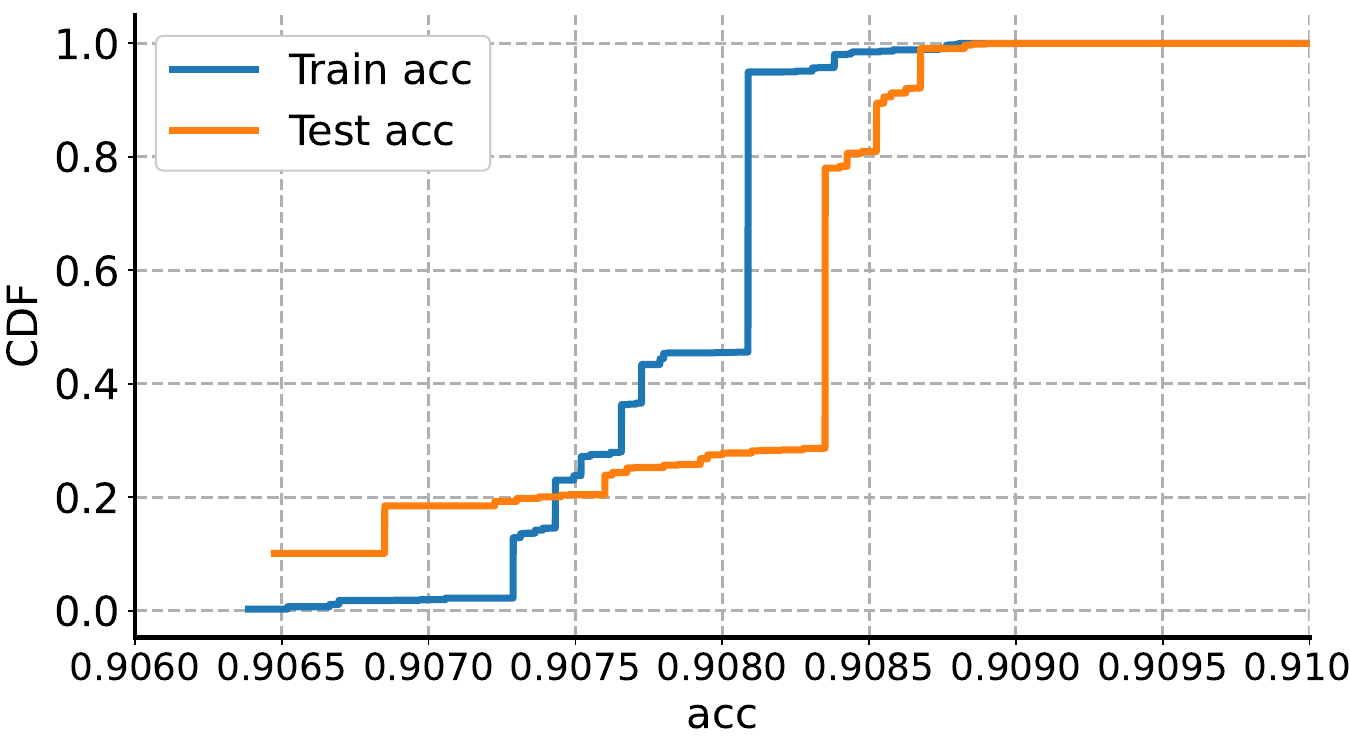}
        \label{fig:acc}
        }

         \vspace{-10pt}
         \caption{Characteristics of ABR in the Rashomon Set}
         \vspace{-15pt}
         \label{fig:Attributes}
\end{figure}
\begin{figure}
    \centering
         \subfigure[Norway]{
        \includegraphics[width=0.46\linewidth]{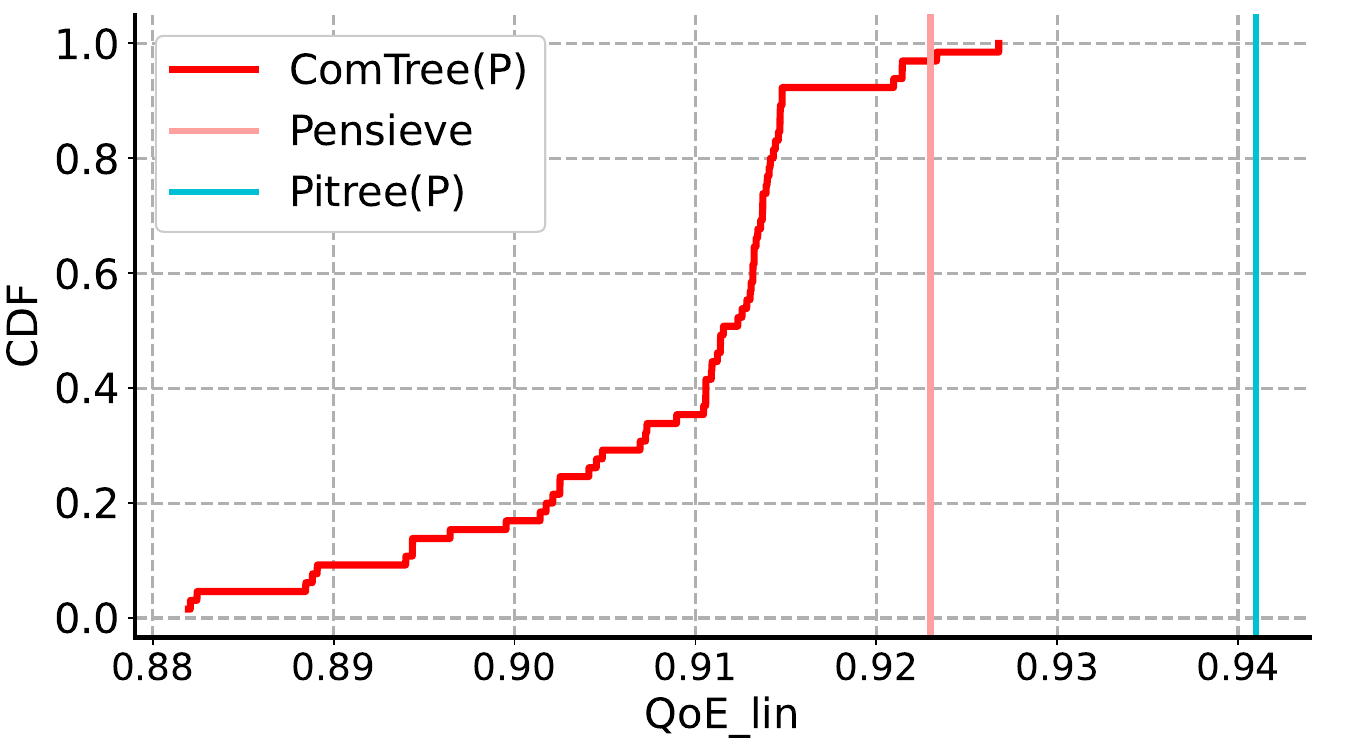}
        \label{fig:rsetn}
        }
            \subfigure[Oboe]{
        \includegraphics[width=0.46\linewidth]{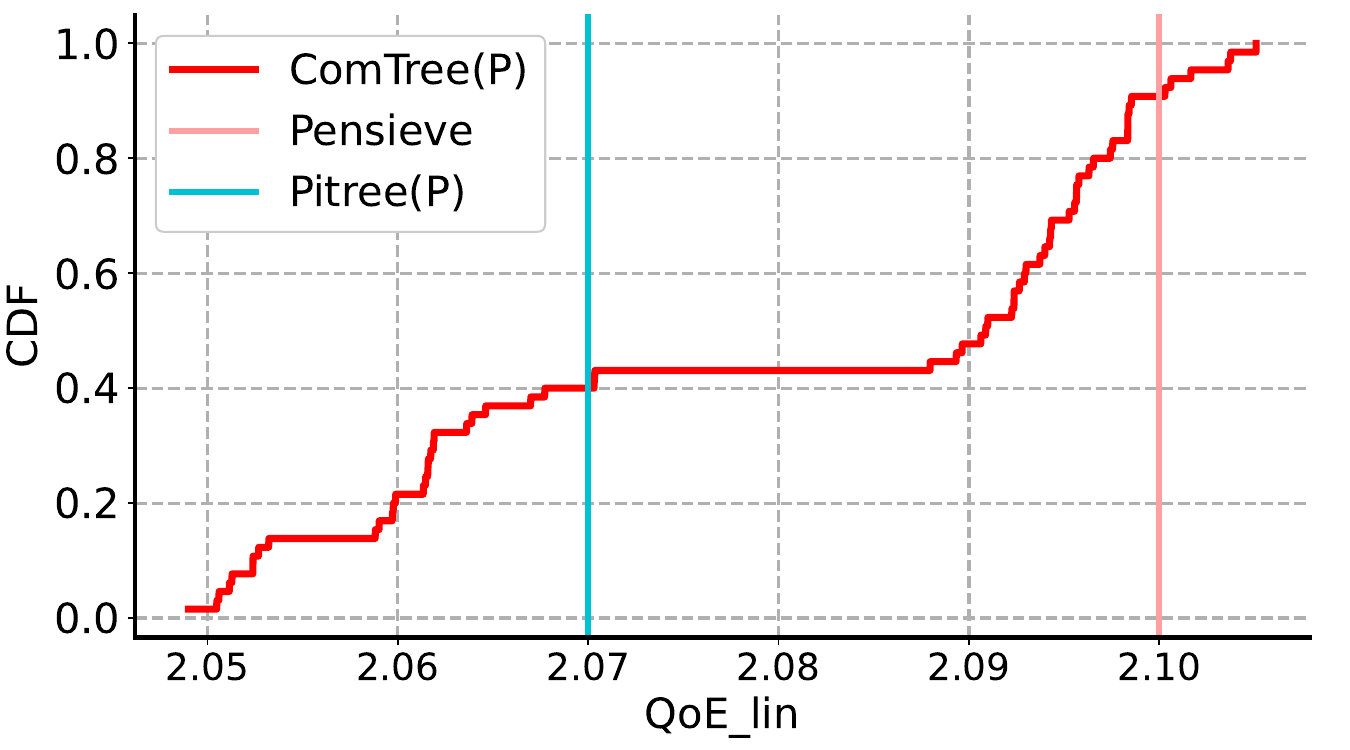}
        \label{fig:rseto}
        }
        \subfigure[Puffer-Oct.17-21]{
        \includegraphics[width=0.46\linewidth]{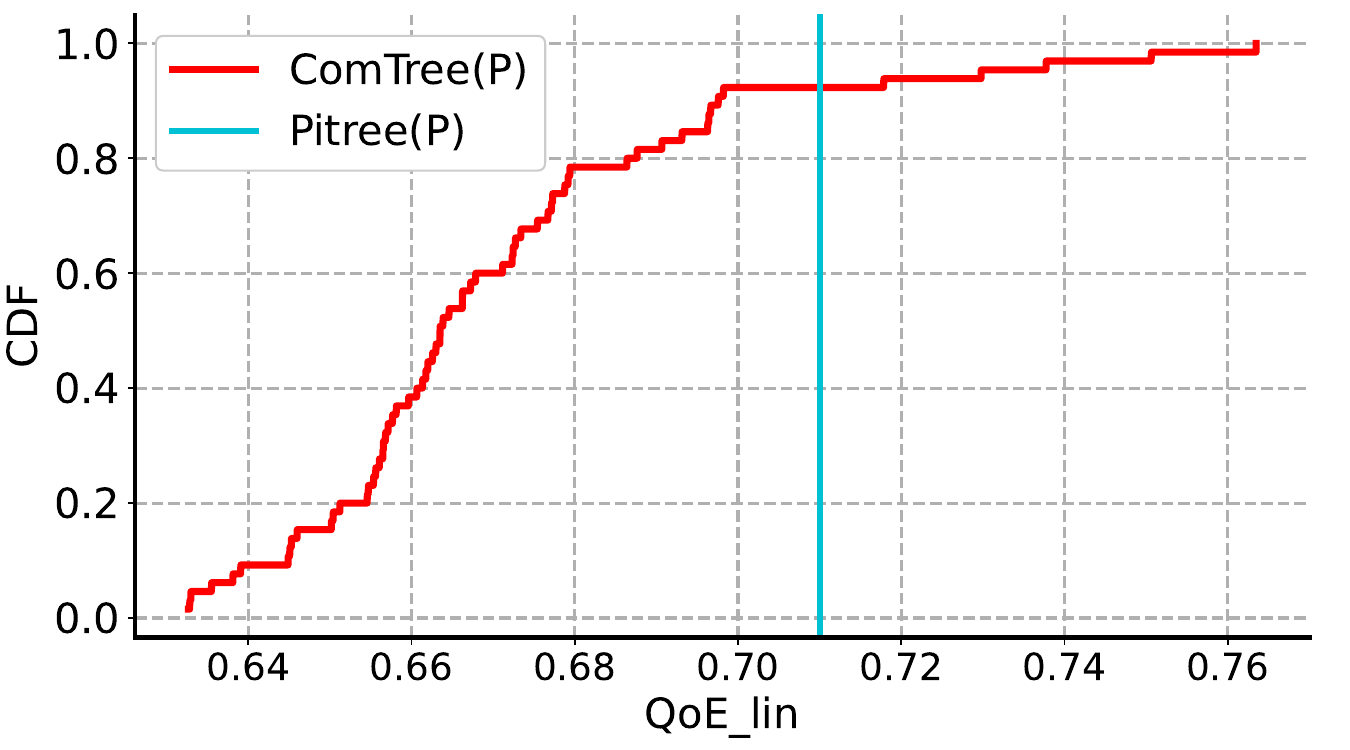}
        \label{fig:rsetp1}
        }
        \subfigure[Puffer-Feb.18-22]{
        \includegraphics[width=0.46\linewidth]{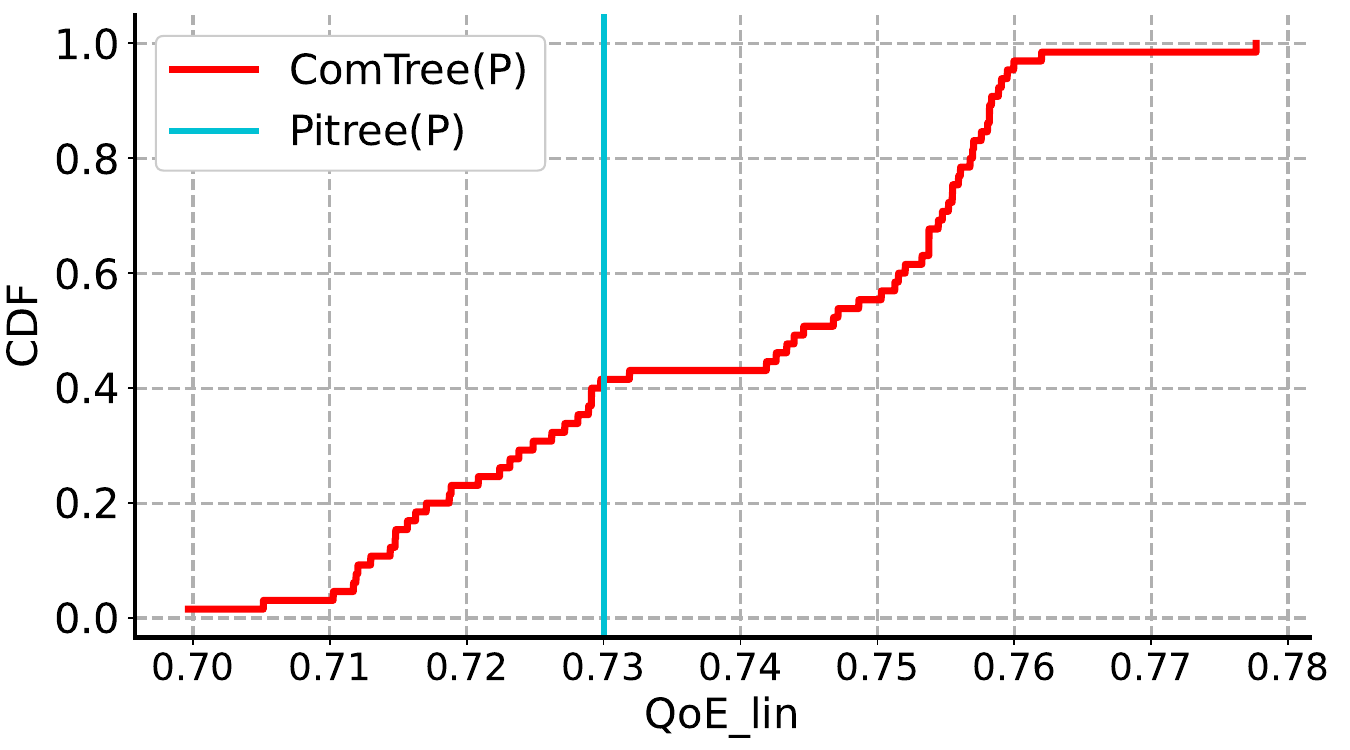}
        \label{fig:rsetp2}
        }
         \vspace{-10pt}
         \caption{$QoE_{lin}$ of ABR in the Rashomon Set}
         \vspace{-15pt}
         \label{fig:QoE in rset}
\end{figure}

\subsubsection{Characteristics of the Rashomon Set}
The analysis covers 64 instances containing 1.6e5 trees, arranged in ascending order based on the Rashomon set's objective values (each instance represents a distinct tree partition). Figure~\ref{fig:number} shows the distribution of decision tree quantities across different instances. The number of decision trees varies significantly among instances, where 10\% of instances contain over 5,000 trees. Figure~\ref{fig:feature} demonstrates the feature utilization distribution (frequency of feature occurrences divided by total tree counts). Feature utilization rates show substantial variation. Nearly 20\% of features appear in less than 1\% of trees, while another 20\% appear multiple times within individual trees on average. Figure~\ref{fig:leaf} displays the leaf node counts distribution, and Figure~\ref{fig:acc} presents the accuracy rates on both training and testing sets. Both metrics demonstrate minimal variation. Over 95\% of decision trees contain between 19 and 21 leaves, with accuracy rates ranging from 0.906 to 0.909 across all trees.

\subsubsection{$QoE_{lin}$ in the Rashomon Set}
While all decision trees within the Rashomon set achieve accuracy within the specified range, as an ABR algorithm, the primary focus lies on video playback QoE rather than accuracy. The evaluation encompasses all instances in the Rashomon set using four network datasets in the simulated environment. Figure~\ref{fig:QoE in rset} illustrates the $QoE_{lin}$ distribution, highlighting the positions of both the teacher algorithm (Pensieve) and the baseline algorithm (PiTree(P)).

The $QoE_{lin}$ variance exhibits dataset-dependent characteristics. The maximum deviation reaches 0.12 in the Puffer-Oct. 17-21 dataset, while the Norway dataset shows the minimal deviation of 0.06. Nevertheless, even the lowest-performing instances remain competitive compared to alternative algorithms. In the Puffer-Oct. 17-21 dataset, where the deviation peaks, the worst-performing instance from the Rashomon set still outperforms all algorithms except RobustMPC and PiTree(P). Using PiTree(P) as a reference baseline, over 60\% of instances surpass its performance on both the Oboe and Puffer-Feb. 18-22 datasets. Thus, the entire Rashomon set demonstrates competitive QoE performance. Moreover, considering the selected Rashomon set encompasses 64 instances and 1.6e5 trees, superior performance can be achieved by reducing the selected instances.




\begin{figure}
    \centering
       \subfigure[Number of Winners per Round]{
        \includegraphics[width=0.46\linewidth]{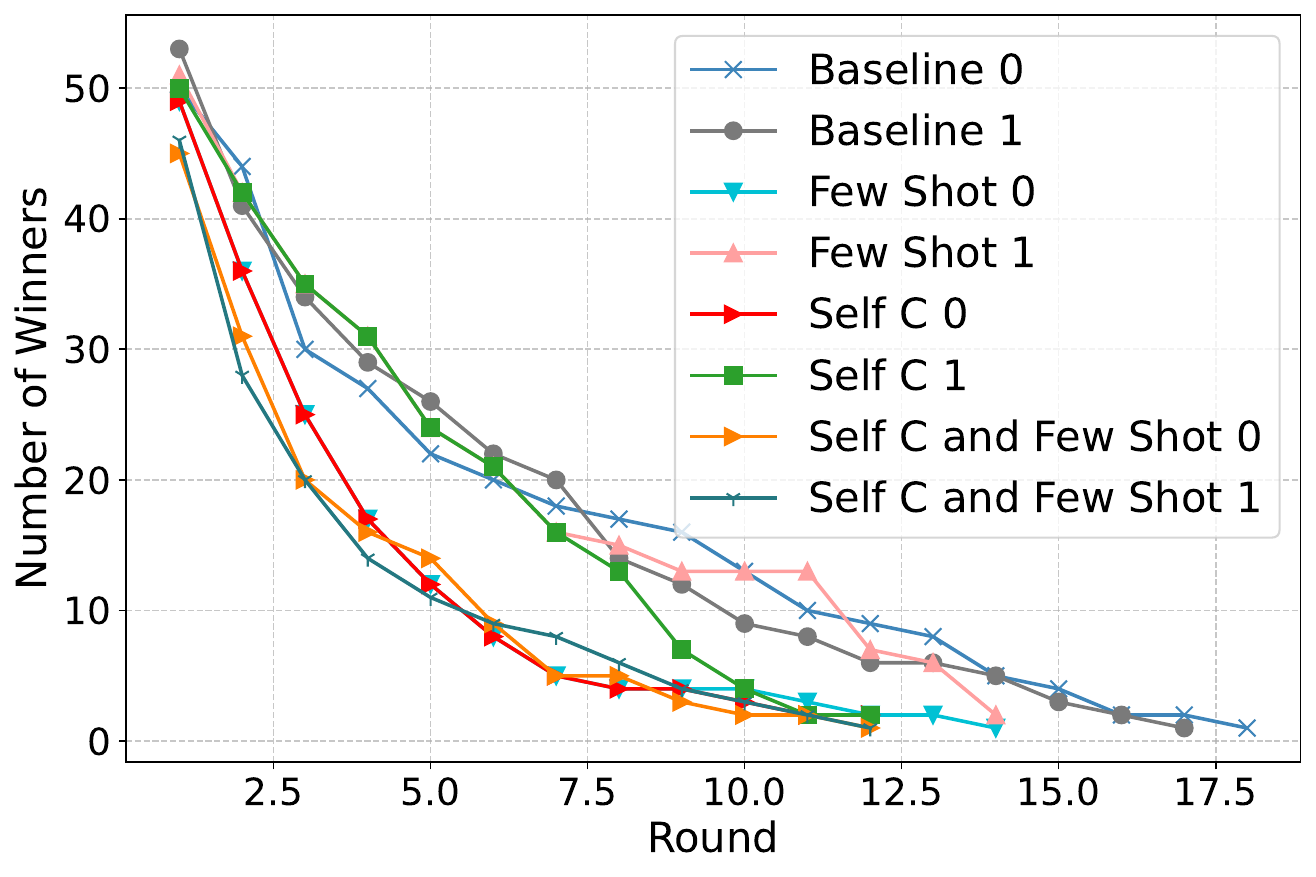}
        \label{fig:winner_counts}
        }
         \subfigure[Ratio of Two Winner   per Round]{
        \includegraphics[width=0.46\linewidth]{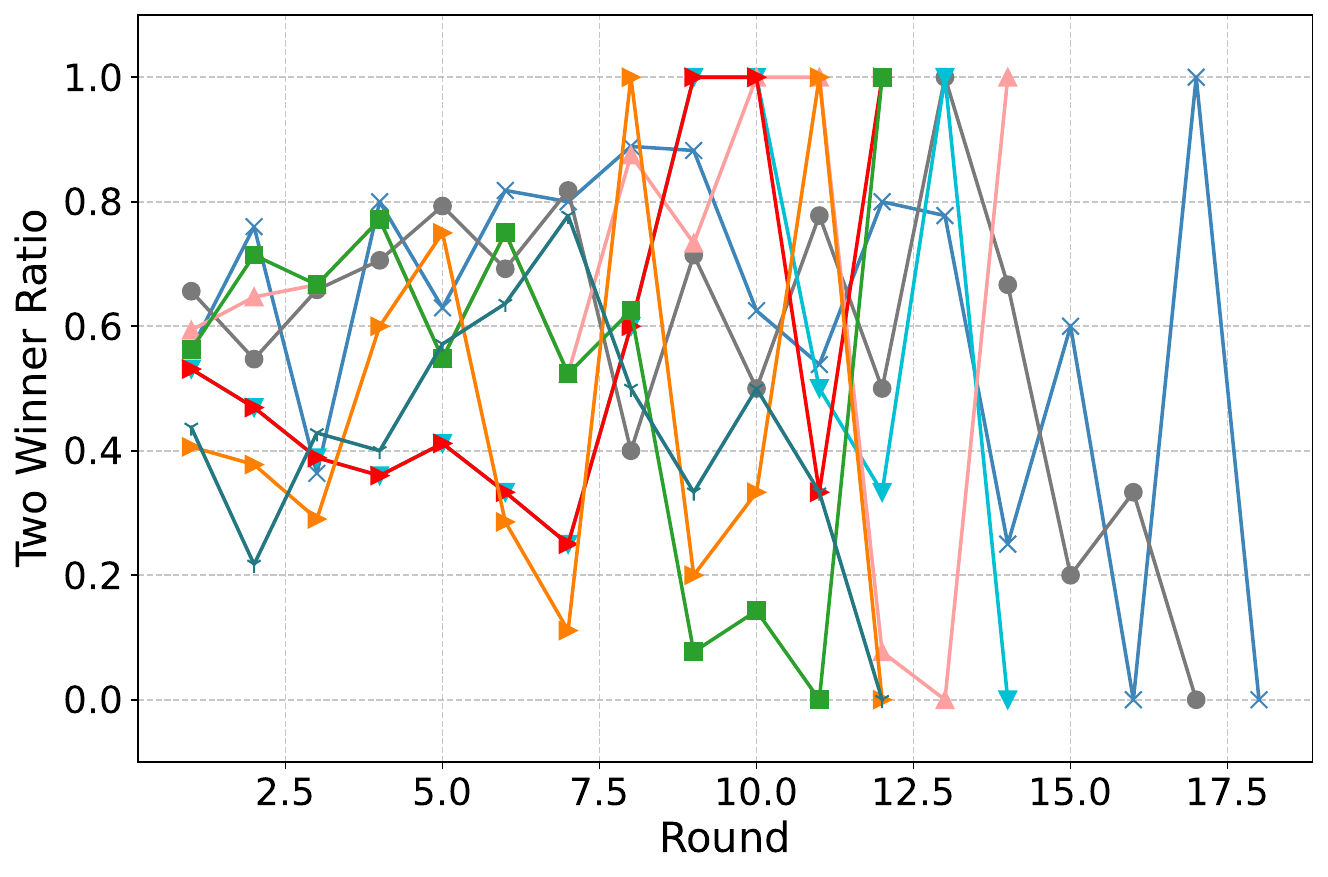}
        \label{fig:winner_idx_ratios}
        }

         \vspace{-5pt}
         \caption{Results of Different Rounds}
         \vspace{-15pt}
         \label{fig:Different Rounds}
\end{figure}

\begin{figure}[h]
    \centering
         \subfigure[Overlap Ratio of Winners per Round]{
        \includegraphics[width=0.46\linewidth]{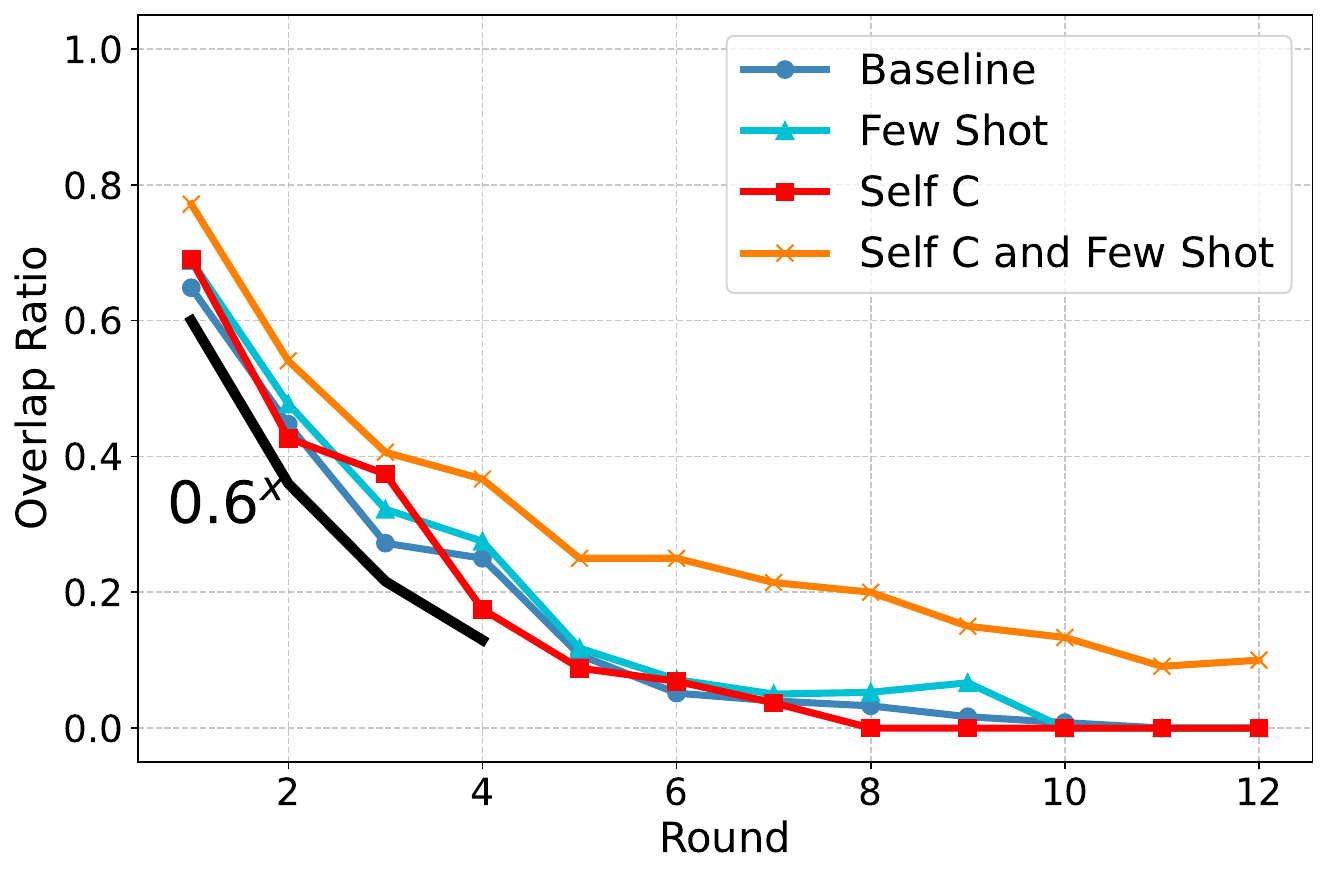}
        \label{fig:overlap_ratios}
        }
        \subfigure[CDF of Ranking Differences]{
        \includegraphics[width=0.46\linewidth]{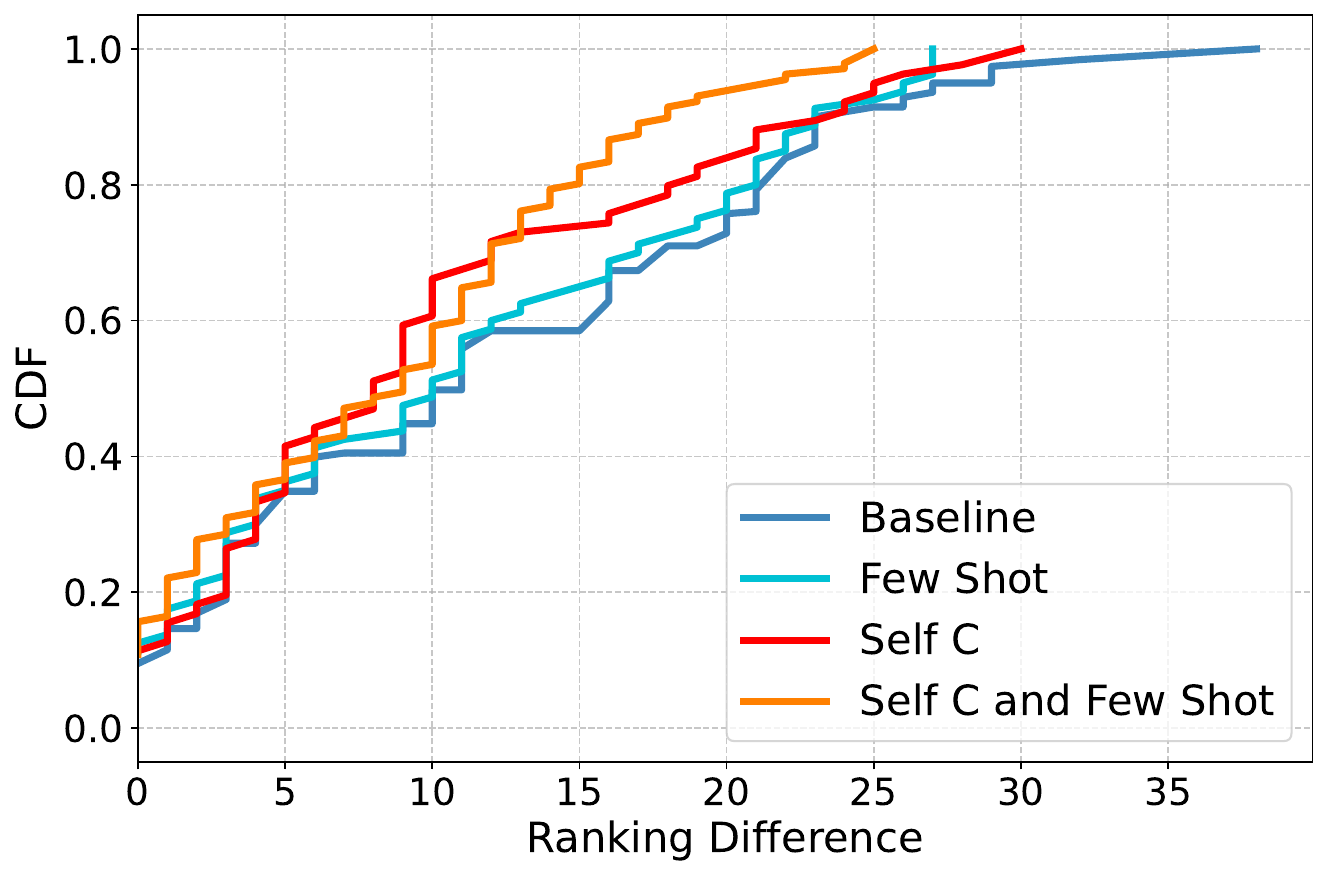}
        \label{fig:ranking_differences}
        }
         \vspace{-5pt}
         \caption{Results of Different Runs}
         \vspace{-15pt}
         \label{fig:Different Runs}
\end{figure}
\begin{figure} 
  \centering
  \includegraphics[width=0.99\linewidth]{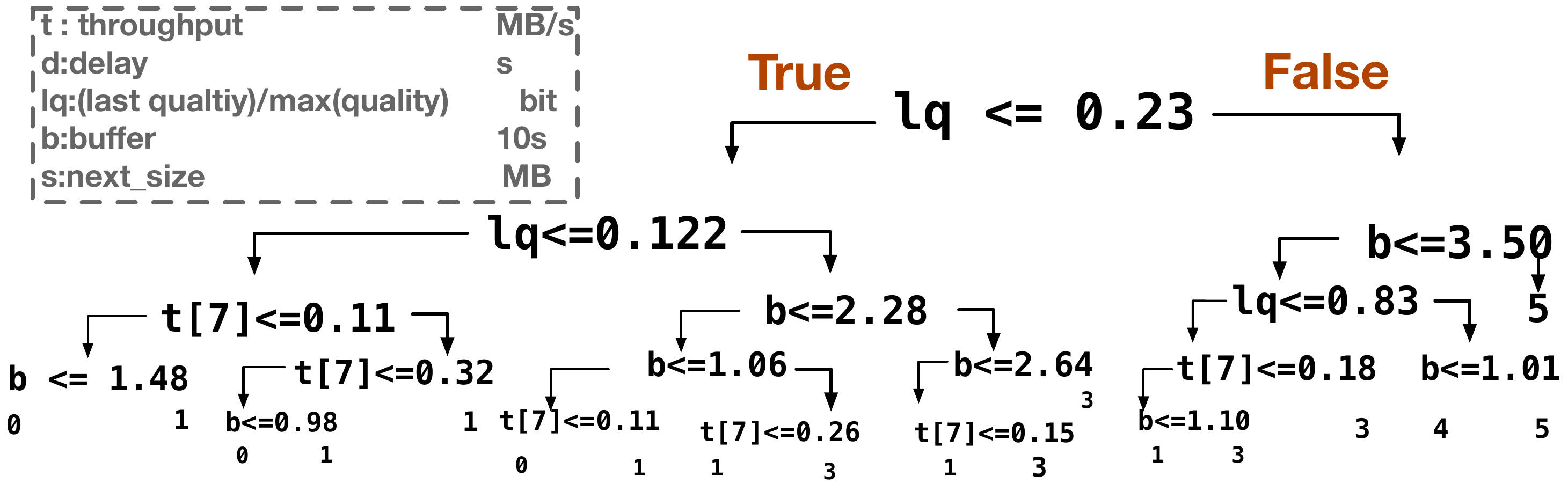}
    \vspace{-10 pt}
  \caption{\texttt{ComTree}\_C~(Some details omitted)
  }
  \vspace{-15 pt}
  \label{fig:ComTreeI}
\end{figure}
\begin{table}[h]
\centering
\caption{Frequency of Bases for Judging Comprehensibility}
\vspace{-10pt}
\begin{tabular}{l|c}

Basis for Judgment & Frequency \\ \hline
Consistency of features within layers or subtrees & 100\% \\ 
Number of layers (depth) in the tree & 100\%  \\ 
Overall structure and organization of the tree & 93.75\% \\ 
Intuitively reasonable feature selection & 93.75\% \\ 
Reasonableness of threshold selection & 20\% \\ 
Complexity of subtrees & 20\%  \\ 
\end{tabular}
\vspace{-10pt}
\label{tab:comprehensibility-bases}
\end{table}

\begin{figure}
    \centering
        \includegraphics[width=0.94\linewidth]{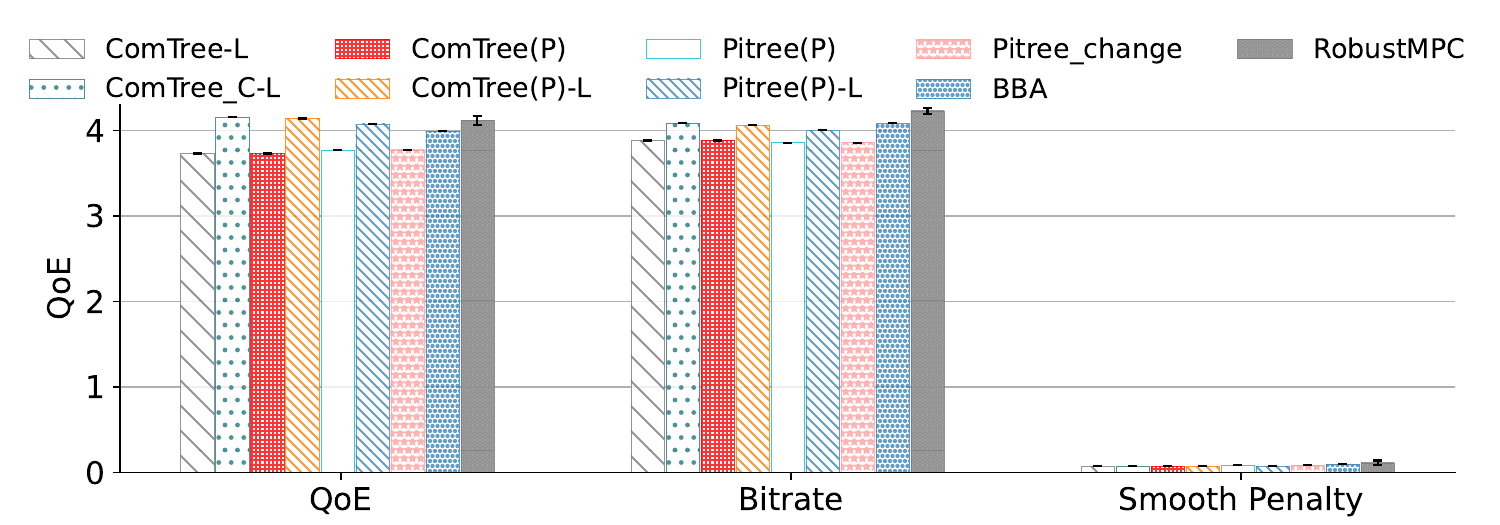}

         \vspace{-10pt}
         \caption{$QoE_{lin}$ after LLM Optimization in 5G}
         \vspace{-15pt}
         \label{fig:after LLM Optimization}
\end{figure}

\subsection{Comprehensibility Analysis of \texttt{ComTree}}\label{sec:eva_inter}
\subsubsection{Process Analysis of Comprehensibility Assessment}
The input-output format for large language models is constructed through instruction prompting. The decision trees represented in JSON format are converted to Python code. LLMs first output their preference, followed by their reasoning. Two prompting engineering methods, $Few-Shot$~\cite{brown2020gpt3} and $Self-Consistency$~\cite{wang2022consistency}, are employed, resulting in four combinations. The details of prompt engineering can be found in Appendix~\ref{app:prompt}.
Figure~\ref{fig:Different Rounds} illustrates the results between different rounds within the same run. Figure~\ref{fig:winner_counts} shows the number of remaining decision trees after each round. These eight experiments require between 12 to 18 rounds. Both $Self-Consistency$ and $Few-Shot$ methods reduce the required rounds from 18 to 12-14. When two large language models disagree, both elements are considered winners. Figure~\ref{fig:winner_idx_ratios} displays the proportion of two winners in each round. The proportion of inconsistent judgments increases with rounds. In four tests, the large language models ultimately select a set of decision trees that cannot be further evaluated.

The consistency between different runs is evaluated in Figure~\ref{fig:Different Runs}. Figure~\ref{fig:overlap_ratios} shows the overlap ratio of winners in each round between two runs, with curve $y=0.6^x$ as reference. The curves consistently exceed the reference curve, indicating over 60\% overlap in each round. Figure~\ref{fig:ranking_differences} presents the distribution of ranking differences for each instance between two runs. For 50\% of the instances, the final ranking difference is within 10, and when using the $Self-Consistency$ method, 40\% of the instances show ranking differences within 5.

\subsubsection{Decision Basis Analysis of LLMs Assessment}
To demonstrate the rationale behind different selections by LLMs, we analyze their responses. We extract responses from the first 32 rounds, totaling 296k words, and analyze the basis for comprehensibility judgments, as shown in Table~\ref{tab:comprehensibility-bases}. The superior comprehensibilityof \texttt{ComTree}\_C~(Figure~\ref{fig:ComTreeI}) compared to \texttt{ComTree}(P) and Pitree(P) (Figure~\ref{fig:pitree}) is evident in the following aspects:
\begin{itemize}[leftmargin=8pt]
    \item Regarding \textbf{tree size} (code lines, leaf nodes, tree depth), Pitree(P) $>>$ \texttt{ComTree}(P) $>$ \texttt{ComTree}\_C
    \item In terms of \textbf{feature count}, Pitree(P) utilizes 14 different features, while \texttt{ComTree}(P) uses 4, and \texttt{ComTree}\_C employs only 3 features. \texttt{ComTree}\_C considers buffer, last quality, and last throughput, the three most crucial features in adaptive bitrate algorithms.
    \item For \textbf{node organization}, compared to \texttt{ComTree}(P), \texttt{ComTree}\_C tends to use one feature multiple times in initial splits with less feature crossing. 
\end{itemize}

\subsection{The Potential for Comprehensibility in \texttt{ComTree}}\label{sec:eva_potential}
To further validate the potential of \texttt{ComTree} for developer enhancement, we design a reproducible experiment simulating adaptation to new network environments. Instead of relying on traditional manual tuning by developers, we adopt an LLM-based approach to modify the algorithms, thereby eliminating human bias in algorithm adjustment. We utilize 5G network traces~\cite{narayanan2020lumos5g} as the testing environment, which exhibits significantly different distributions from previous training and testing traces. We provide the LLM with fundamental ABR knowledge through prompts, including inputs, outputs, and optimization objectives, along with one network trace as optimization context, enabling the LLM to act as a professional engineer optimizing ABR algorithms. We compare the performance of three different ABR algorithms before and after LLM optimization, including the most comprehensible decision tree \texttt{ComTree}\_C, the optimal tree \texttt{ComTree}(P), and the previous work Pitree(P), with optimized decision trees denoted by the '-L' suffix.

Figure~\ref{fig:after LLM Optimization} illustrates the $QoE_{lin}$ and its components for different algorithms before and after LLM optimization in a 5G network environment. The experimental results highlight the critical role of comprehensibility in algorithm optimization from two perspectives.
First, we observe a strong positive correlation between the comprehensibility of decision trees and their optimization potential. The adjusted performance follows the order \texttt{ComTree}\_C-L > \texttt{ComTree}(P)-L > \texttt{Pitree}(P)-L, which is consistent with the comprehensibility evaluation of \texttt{ComTree}. This demonstrates that higher comprehensibility provides greater optimization potential.
Furthermore, comprehensibility has practical value. After optimization, \texttt{ComTree}\_C-L achieves a $QoE_{lin}$ of 4.16, successfully surpassing the classical algorithm RobustMPC (4.12). This indicates that \texttt{ComTree}, by improving comprehensibility, enables developers to conveniently modify algorithms and tune performance, thereby surpassing pre-designed algorithms.

%% file: discuss.tex



%% file: conclude.tex
\section{Conclusion}
We proposed \texttt{ComTree}, the first bitrate adaptation algorithm generation framework that explored comprehensibility as a novel optimization objective. Our approach generated a Rashomon set containing all decision trees that satisfied accuracy requirements, and subsequently developed a comprehensibility evaluation framework utilizing LLMs. Through extensive experiments, our algorithm demonstrated competitive performance in both performance and comprehensibility, and showcased the potential for \texttt{ComTree} to be further enhanced through developer improvements.

\noindent\textbf{Acknowledgments} 
We thank the anonymous MM reviewers for their constructive feedback, which has significantly improved this work exploring new optimization directions. This research was supported by the Beijing National Research Center for Information Science and Technology under Grant BNR2023TD03005-2, the NSERC Discovery Grant, and the Beijing Key Laboratory of Networked Multimedia.

%% file: app.tex
\appendix
\section{Feature Processing and Rashomon Set Construction Algorithm}~\label{app:alg}
 \begin{algorithm}[h]
    \caption{Feature Processing Algorithm}
    \label{alg:fp}
    \KwIn{state-action dataset $(S, A)$, importance threshold  $\delta$, reference accuracy $acc_{rec}$, 
    
    }
    \KwOut{new dataset $(S_{guess},A)$}

    {\color{teal}// Obtain initial accuracy and column importance } \\
    $acc_{ori}, importance= Xgboost(S,A)$ \\
    $S_{guess} = S.copy$, $acc_{new} = acc_{ori}$ {\color{teal}//init dataset}\\

    {\color{blue}{/*Column elimination while preserving accuracy}} \\
    \While{$acc_{new}  >= min(acc_{ori}, acc_{rec})~ and ~len(S_{guess}.column)>1$}{
        {\color{teal}// Column elimination based on importance threshold} \\
        $S_{guess}= fliter\_importance(S_{guess}, importance, \delta)$ \\
        {\color{teal}// Column elimination based on least important columns} \\
        $c_{min} = min(importance)$ \\
        $S_{guess} = S_{guess}.delete(c_{min})$ \\
        $acc_{new}, importance= Xgboost(S_{guess},A)$\\
    }
    {\color{teal}// Revert column elimination that degrades accuracy} \\
    \If{ $c_{min}!=empty$}{
        $S_{guess} =  S_{guess}.add(c_{min})$
    }
     \Return $(S_{guess},A)$ \\
    
\end{algorithm}
 \begin{algorithm}
    \caption{Rashomon Set Construction Algorithm}
    \label{alg:Rashomon}
    \KwIn{ABR Algorithm $\pi^{*}$, maximum iteration number $M$, 
    regularization parameter $\lambda$, Rashomon set bounds parameter $\epsilon$, max depth $d$}
    \KwOut{Rashomon set $R_{set}$}
    
    {\color{blue}{/*Construct a dataset that conforms to real occurrence probabilities using the teacher-student learning framework}} \\
    $(S,A) =  VirtualPlay(\pi^*)$ {\color{teal}//init dataset}\\
    \For{$i \in [1,...,M]$}{
        {\color{teal}// Reduce feature number via column elimination } \\
        $(S_{guess},A) = Feature\_Processing(S,A)$\\
        {\color{teal}// Train decision tree via the reduced dataset} \\
        $\pi_{i} = TrainDT(S_{guess},A,d)$\\
        {\color{teal}// Update dataset through the teacher-student learning} \\
        $(S_s,A_s) =  VirtualPlay(\pi_{i})$\\
        $A_t = Predict(\pi^*,S_s)$\\
        $(S,A) = (S,A) \bigcup (S_s, A_t)$
    }
    {\color{blue}{/*Generate Rashomon set via TreeFarms}} \\
    $(S_{guess},A) = Feature\_Processing(S,A)$\\
    {\color{teal}// Get the best decision tree via GOSDT algorithm} \\
    $obj_{opt} = GOSDT(S_{guess},A,\lambda,d)$\\
    {\color{teal}// Set he threshold of the Rashomon set} \\
    $\theta_{\epsilon} = obj_{opt} * (1 + \epsilon )$\\
    {\color{teal}// Generate Rashomon set via TreeFarms} \\
    $R_{set} = TreeFarms(S_{guess},A,\lambda,\theta_{\epsilon},d)$\\
     \Return $R_{set}$ \\
\end{algorithm}

\section{Positioning of ComTree: Balancing Performance and Comprehensibility}~\label{app:com}
ABR algorithms, as core algorithms that directly impact user QoE, have performance as their primary evaluation metric. However, whether based on simulation or real-world testing, it is difficult to fully reproduce the infinite variations of scenarios in real networks, making algorithm generalization a critical challenge.

To improve generalization, researchers have proposed black-box methods based on meta-reinforcement learning. For example, MERINA~\cite{kan2022merina} achieves performance superior to Pensieve on datasets such as Puffer by modeling environmental variables. Despite their superior performance, the ``black-box'' nature of these methods presents inherent challenges: when algorithms perform poorly in unforeseen scenarios (bad cases), developers struggle to debug and optimize effectively due to their inability to understand the internal decision logic. This is a major challenge faced by all black-box methods.

To address this challenge, the academic community has begun focusing on algorithm interpretability, aiming to transform black-box models into white-box models that humans can understand, such as Pitree. However, we argue that for engineering systems like ABR algorithms that require continuous iteration and maintenance, being merely ``interpretable'' is insufficient. Interpretability typically answers \textbf{``How did the model decide?''} but does not necessarily tell developers \textbf{``Why was the model designed this way?''} Therefore, we propose a more advanced concept—\textbf{comprehensibility}—which is fundamentally concerned with ensuring that the algorithm's structure and logic are not only white-box, but also easily understood and conveniently adjusted and improved by developers, thereby endowing the algorithm with the potential for continuous evolution. This represents an important approach to enhancing the algorithm's generalization capability in real-world production environments.

Based on the two key dimensions of performance and comprehensibility, existing ABR algorithms can be categorized as follows:

\begin{itemize}
\item Traditional heuristic algorithms (e.g., BBA, MPC): These algorithms are based on fixed rules and possess high comprehensibility, but due to rigid rules, they struggle to adapt to complex network environments, resulting in generally not good QoE performance.
\item Black-box methods (e.g., Pensieve, MERINA): These algorithms typically achieve top-tier QoE performance, but their complex internal structures (such as deep neural networks) result in low comprehensibility, leaving developers helpless when facing bad cases.
\item Hybrid methods (e.g., Oboe~\cite{akhtar2018oboe}): These methods attempt to find a balance between performance and comprehensibility. For example, Oboe guides decisions through parameterized rule tables, which can be seen as an attempt to sacrifice comprehensibility to some extent for improved performance.
\end{itemize}

In this context, ComTree is positioned as the first ABR algorithm to provide high comprehensibility while always maintaining SOTA-level performance. As shown in the experimental results in Table~\ref{table:app}, ComTree's performance closely follows its ``teacher'' models (such as MERINA) and has the potential to improve as the ``teacher'' advances. This is due to its unique tree structure, which can effectively distill knowledge from black-box models while maintaining logical clarity for easy analysis and intervention by developers. In contrast, methods like Oboe have limited performance improvement potential due to their fixed structures. Therefore, ComTree is not only competitive in current performance but, more importantly, provides a new path for continuous iteration and maintenance of ABR algorithms that balances both performance and comprehensibility.

In summary, this paper not only proposes a new algorithm called ComTree but, more importantly, introduces and explores ``comprehensibility'' as a new evaluation dimension. We believe this metric will open new directions for future ABR algorithm research and drive the community to build next-generation video transmission systems that are both powerful and reliable.
\begin{table}[h]
\centering
\caption{Performance comparison with parameter adjustment algorithms and meta-learning algorithms}
\begin{tabular}{l|ccc}
\textbf{Dataset} & \textbf{MERINA} & \textbf{ComTree(M)} & \textbf{Oboe} \\
\hline
Norway & 0.97($\pm$ 0.61)  & 0.94($\pm$ 0.65) & 0.88($\pm$ 0.68) \\
\hline
Oboe & 2.20($\pm$ 1.09)  & 2.19($\pm$ 1.10) & 2.19($\pm$ 1.15) \\
\hline
Puffer-2110 & 0.77($\pm$ 2.01) & 0.76($\pm$ 1.98) & 0.77($\pm$ 1.89) \\
\hline
Puffer-2202 & 0.85($\pm$ 2.98) & 0.82($\pm$ 2.99) & 0.77($\pm$ 3.10) \\
\label{table:app}
\end{tabular}
\end{table}

\section{Limitations and Future Work}~\label{app:dis}
\textbf{Is ComTree the optimal solution for comprehensibility?}
Although \texttt{ComTree} achieves competitive performance on multiple datasets using only 2/3 of the nodes of Pitree, and shows a certain level of stability in the evaluation by LLMs, there is still a gap between \texttt{ComTree} and the optimal comprehensibility solution. The main reason lies in feature selection. According to Table~\ref{tab:comprehensibility-bases}, feature selection and threshold selection are among the top 4 factors influencing comprehensibility. However, to reduce computational overhead, \texttt{ComTree} borrows knowledge from black-box algorithms for feature partitioning. Balancing computational overhead and the comprehensibility of feature selection is the key to developing more advanced comprehensibility algorithms.

\textbf{Where are the capability boundaries of LLMs?}
Referencing the conclusions from other fields~\cite{ahn2022can,driess2023palm}, LLMs are more suitable for macro-level planning and content understanding rather than specific decision-making. Therefore, we employ LLMs to evaluate comprehensibility. Although we adopt a series of methods to enhance the reliability of LLMs' comprehensibility assessments and LLMs have demonstrated near-human subjective understanding capabilities in many fields~\cite{park2023generative,aher2023using,hussain2024tutorial,ren2024bases,zhao2023recommender,huang2023recommender}, the specific gap between LLMs' evaluations and human subjective perception remains an open question. This requires large-scale subjective experiments with domain-knowledgeable experts, which is not only a demand in streaming media but also a challenge for the entire AI community to address.

\section{Prompt Engineering in Comprehensibility Assessment and Adjustment}~\label{app:prompt}
We show the prompt used in \S\ref{sec:eva_rset} for comparing the comprehensibility as Figure~\ref{prompt:interpretability}. We input two decision trees converted to Python code and then specify the output format with the first word being the preference followed by the reason. 
To improve the stability of the LLMs' performance, we also consider $Few-Shot$ learning and $Self-Consistency$ prompt engineering methods. The specific designs are as follows:
\begin{itemize}
    \item $Few-Shot$: Before evaluating the LLMs, we first introduce the LLMs of some prior knowledge, which in our task is how developers understand the decision trees converted from ABR. Specifically, we specify which aspects are easier for developers to understand in terms of the number of leaves, tree depth, feature importance, and the organization of the tree. Our specific prompt can be seen in Figure~\ref{prompt:few-shot}.
    \item $Self-Consistency$: In each round, we repeat the query to each LLM 3 times and take the preference that appears more frequently as the final answer of that LLM for that round.
\end{itemize}
We present the prompt for LLM adjustment used in \S\ref{sec:eva_potential} in Figure~\ref{prompt:llma}, where one 5G trace serves as the knowledge base, and the prompt defines background information including input, output, and optimization objectives.

\begin{figure*}[!htb]
\begin{promptbox}
\begin{verbatim}
Which of the following two decision trees has higher comprehensibility?
Please use \'TREEONE\' as the first word in your answer to indicate that the first decision tree has 
higher comprehensibility, or use \'TREETWO\' to indicate that the second decision tree has higher comprehensibility,
followed by an explanation.
Decision tree 1: {0}.Decision tree 2: {1}.'.format(tree1,tree2)
\end{verbatim}
\end{promptbox}
\caption{Prompt for Comparing the Comprehensibility of Two Decision Trees}
\label{prompt:interpretability}
\end{figure*}

\begin{figure*}[!htb]
\begin{promptbox}
\begin{verbatim}
As a developer in the field of streaming media, 
we typically assess the comprehensibility of a decision tree for bitrate adaptation algorithms in the following ways.
First and foremost, we consider the number of layers in the tree; 
the fewer the layers, the stronger the comprehensibility. 
Secondly, we prefer features that are more intuitive. Generally speaking, we desire important features 
such as last_quality, buffer, and tput. 
Furthermore, we aspire for the decision tree to be well-organized, with features within the same layer 
or subtree being as consistent as possible. 
The aforementioned methods are merely examples of evaluating comprehensibility. 
In practice, comprehensibility often implies better comprehension for developers
\end{verbatim}

\end{promptbox}

\caption{Prompt for $Few-Shot$}
\label{prompt:few-shot}
\end{figure*}
\begin{figure*}[!htb]
\begin{promptbox}
\begin{verbatim}
You are a streaming media expert.
I hope you can adjust the input decision tree, which is a bitrate adaptation algorithm.
It was designed for an environment with average bandwidth of 1.31 ± 0.69Mbps, range [0.13, 3.42]mbps.
I want it to perform well in an environment with average bandwidth of 344.68 ± 127.63Mbps, range [64.10, 564.12]mbps.
The variables in the decision tree have units of :
throughput (Mbyte/s), curr_buffer (10 seconds), last_quality=video[last_act]/4300,
video=[300, 750, 1200, 1850, 2850, 4300] in kbps,
and the final choice is the video bitrate.
I want video[act]/1000-4.3*rebuf-abs(video[act]-video[last_act])/1000.0 to be as large as possible.
Please output the modified decision tree directly in the form of Python code.
\end{verbatim}

\end{promptbox}
\caption{Prompt for LLM Adjustment}
\label{prompt:llma}
\end{figure*}

%% file: main.bbl

\begin{thebibliography}{61}


\ifx \showCODEN    \undefined \def \showCODEN     #1{\unskip}     \fi
\ifx \showDOI      \undefined \def \showDOI       #1{#1}\fi
\ifx \showISBNx    \undefined \def \showISBNx     #1{\unskip}     \fi
\ifx \showISBNxiii \undefined \def \showISBNxiii  #1{\unskip}     \fi
\ifx \showISSN     \undefined \def \showISSN      #1{\unskip}     \fi
\ifx \showLCCN     \undefined \def \showLCCN      #1{\unskip}     \fi
\ifx \shownote     \undefined \def \shownote      #1{#1}          \fi
\ifx \showarticletitle \undefined \def \showarticletitle #1{#1}   \fi
\ifx \showURL      \undefined \def \showURL       {\relax}        \fi
\providecommand\bibfield[2]{#2}
\providecommand\bibinfo[2]{#2}
\providecommand\natexlab[1]{#1}
\providecommand\showeprint[2][]{arXiv:#2}

\bibitem[Achiam et~al\mbox{.}(2023)]%
        {achiam2023gpt}
\bibfield{author}{\bibinfo{person}{Josh Achiam}, \bibinfo{person}{Steven Adler}, \bibinfo{person}{Sandhini Agarwal}, \bibinfo{person}{Lama Ahmad}, \bibinfo{person}{Ilge Akkaya}, \bibinfo{person}{Florencia~Leoni Aleman}, \bibinfo{person}{Diogo Almeida}, \bibinfo{person}{Janko Altenschmidt}, \bibinfo{person}{Sam Altman}, \bibinfo{person}{Shyamal Anadkat}, {et~al\mbox{.}}} \bibinfo{year}{2023}\natexlab{}.
\newblock \showarticletitle{Gpt-4 technical report}.
\newblock \bibinfo{journal}{\emph{arXiv preprint arXiv:2303.08774}} (\bibinfo{year}{2023}).
\newblock


\bibitem[Aher et~al\mbox{.}(2023)]%
        {aher2023using}
\bibfield{author}{\bibinfo{person}{Gati~V Aher}, \bibinfo{person}{Rosa~I Arriaga}, {and} \bibinfo{person}{Adam~Tauman Kalai}.} \bibinfo{year}{2023}\natexlab{}.
\newblock \showarticletitle{Using large language models to simulate multiple humans and replicate human subject studies}. In \bibinfo{booktitle}{\emph{International Conference on Machine Learning}}. PMLR, \bibinfo{pages}{337--371}.
\newblock


\bibitem[Ahn et~al\mbox{.}(2022)]%
        {ahn2022can}
\bibfield{author}{\bibinfo{person}{Michael Ahn}, \bibinfo{person}{Anthony Brohan}, \bibinfo{person}{Noah Brown}, \bibinfo{person}{Yevgen Chebotar}, \bibinfo{person}{Omar Cortes}, \bibinfo{person}{Byron David}, \bibinfo{person}{Chelsea Finn}, \bibinfo{person}{Chuyuan Fu}, \bibinfo{person}{Keerthana Gopalakrishnan}, \bibinfo{person}{Karol Hausman}, {et~al\mbox{.}}} \bibinfo{year}{2022}\natexlab{}.
\newblock \showarticletitle{Do as i can, not as i say: Grounding language in robotic affordances}.
\newblock \bibinfo{journal}{\emph{arXiv preprint arXiv:2204.01691}} (\bibinfo{year}{2022}).
\newblock


\bibitem[Akhtar et~al\mbox{.}(2018)]%
        {akhtar2018oboe}
\bibfield{author}{\bibinfo{person}{Zahaib Akhtar}, \bibinfo{person}{Yun~Seong Nam}, \bibinfo{person}{Ramesh Govindan}, \bibinfo{person}{Sanjay Rao}, \bibinfo{person}{Jessica Chen}, \bibinfo{person}{Ethan Katz-Bassett}, \bibinfo{person}{Bruno Ribeiro}, \bibinfo{person}{Jibin Zhan}, {and} \bibinfo{person}{Hui Zhang}.} \bibinfo{year}{2018}\natexlab{}.
\newblock \showarticletitle{Oboe: Auto-tuning video ABR algorithms to network conditions}. In \bibinfo{booktitle}{\emph{Proceedings of the 2018 Conference of the ACM Special Interest Group on Data Communication}}. \bibinfo{pages}{44--58}.
\newblock


\bibitem[Anthropic(2025)]%
        {Claude}
\bibfield{author}{\bibinfo{person}{Anthropic}.} \bibinfo{year}{2025}\natexlab{}.
\newblock \bibinfo{booktitle}{\emph{Claude}}.
\newblock
\urldef\tempurl%
\url{https://www.anthropic.com/claude}
\showURL{%
\tempurl}


\bibitem[Bastani et~al\mbox{.}(2018)]%
        {bastani2018student-teacher}
\bibfield{author}{\bibinfo{person}{Osbert Bastani}, \bibinfo{person}{Yewen Pu}, {and} \bibinfo{person}{Armando Solar-Lezama}.} \bibinfo{year}{2018}\natexlab{}.
\newblock \showarticletitle{Verifiable reinforcement learning via policy extraction}.
\newblock \bibinfo{journal}{\emph{Advances in neural information processing systems}}  \bibinfo{volume}{31} (\bibinfo{year}{2018}).
\newblock


\bibitem[Brown et~al\mbox{.}(2020)]%
        {brown2020gpt3}
\bibfield{author}{\bibinfo{person}{Tom Brown}, \bibinfo{person}{Benjamin Mann}, \bibinfo{person}{Nick Ryder}, \bibinfo{person}{Melanie Subbiah}, \bibinfo{person}{Jared~D Kaplan}, \bibinfo{person}{Prafulla Dhariwal}, \bibinfo{person}{Arvind Neelakantan}, \bibinfo{person}{Pranav Shyam}, \bibinfo{person}{Girish Sastry}, \bibinfo{person}{Amanda Askell}, {et~al\mbox{.}}} \bibinfo{year}{2020}\natexlab{}.
\newblock \showarticletitle{Language models are few-shot learners}.
\newblock \bibinfo{journal}{\emph{Advances in neural information processing systems}}  \bibinfo{volume}{33} (\bibinfo{year}{2020}), \bibinfo{pages}{1877--1901}.
\newblock


\bibitem[Chen and Guestrin(2016)]%
        {chen2016xgboost}
\bibfield{author}{\bibinfo{person}{Tianqi Chen} {and} \bibinfo{person}{Carlos Guestrin}.} \bibinfo{year}{2016}\natexlab{}.
\newblock \showarticletitle{Xgboost: A scalable tree boosting system}. In \bibinfo{booktitle}{\emph{Proceedings of the 22nd acm sigkdd international conference on knowledge discovery and data mining}}. \bibinfo{pages}{785--794}.
\newblock


\bibitem[Cormen et~al\mbox{.}(2022)]%
        {cormen2022introduction}
\bibfield{author}{\bibinfo{person}{Thomas~H Cormen}, \bibinfo{person}{Charles~E Leiserson}, \bibinfo{person}{Ronald~L Rivest}, {and} \bibinfo{person}{Clifford Stein}.} \bibinfo{year}{2022}\natexlab{}.
\newblock \bibinfo{booktitle}{\emph{Introduction to algorithms}}.
\newblock \bibinfo{publisher}{MIT press}.
\newblock


\bibitem[{DASH Industry Forum}(2024)]%
        {dashjs}
\bibfield{author}{\bibinfo{person}{{DASH Industry Forum}}.} \bibinfo{year}{2024}\natexlab{}.
\newblock \bibinfo{title}{dash.js}.
\newblock \bibinfo{howpublished}{\url{https://dashjs.org/}}.
\newblock
\urldef\tempurl%
\url{https://dashjs.org/}
\showURL{%
\tempurl}
\newblock
\shownote{[Online; accessed 15-February-2025]}.


\bibitem[Dethise et~al\mbox{.}(2019)]%
        {dethise2019cracking}
\bibfield{author}{\bibinfo{person}{Arnaud Dethise}, \bibinfo{person}{Marco Canini}, {and} \bibinfo{person}{Srikanth Kandula}.} \bibinfo{year}{2019}\natexlab{}.
\newblock \showarticletitle{Cracking open the black box: What observations can tell us about reinforcement learning agents}. In \bibinfo{booktitle}{\emph{Proceedings of the 2019 Workshop on Network Meets AI \& ML}}. \bibinfo{pages}{29--36}.
\newblock


\bibitem[Devlin et~al\mbox{.}(2019)]%
        {devlin2019bert}
\bibfield{author}{\bibinfo{person}{Jacob Devlin}, \bibinfo{person}{Ming-Wei Chang}, \bibinfo{person}{Kenton Lee}, {and} \bibinfo{person}{Kristina Toutanova}.} \bibinfo{year}{2019}\natexlab{}.
\newblock \showarticletitle{Bert: Pre-training of deep bidirectional transformers for language understanding}. In \bibinfo{booktitle}{\emph{Proceedings of the 2019 conference of the North American chapter of the association for computational linguistics: human language technologies, volume 1 (long and short papers)}}. \bibinfo{pages}{4171--4186}.
\newblock


\bibitem[Driess et~al\mbox{.}(2023)]%
        {driess2023palm}
\bibfield{author}{\bibinfo{person}{Danny Driess}, \bibinfo{person}{Fei Xia}, \bibinfo{person}{Mehdi~SM Sajjadi}, \bibinfo{person}{Corey Lynch}, \bibinfo{person}{Aakanksha Chowdhery}, \bibinfo{person}{Brian Ichter}, \bibinfo{person}{Ayzaan Wahid}, \bibinfo{person}{Jonathan Tompson}, \bibinfo{person}{Quan Vuong}, \bibinfo{person}{Tianhe Yu}, {et~al\mbox{.}}} \bibinfo{year}{2023}\natexlab{}.
\newblock \showarticletitle{Palm-e: An embodied multimodal language model}.
\newblock \bibinfo{journal}{\emph{arXiv preprint arXiv:2303.03378}} (\bibinfo{year}{2023}).
\newblock


\bibitem[Fisher et~al\mbox{.}(2019)]%
        {fisher2019Rashomon2}
\bibfield{author}{\bibinfo{person}{Aaron Fisher}, \bibinfo{person}{Cynthia Rudin}, {and} \bibinfo{person}{Francesca Dominici}.} \bibinfo{year}{2019}\natexlab{}.
\newblock \showarticletitle{All models are wrong, but many are useful: Learning a variable's importance by studying an entire class of prediction models simultaneously}.
\newblock \bibinfo{journal}{\emph{Journal of Machine Learning Research}} \bibinfo{volume}{20}, \bibinfo{number}{177} (\bibinfo{year}{2019}), \bibinfo{pages}{1--81}.
\newblock


\bibitem[Freitas(2006)]%
        {freitas2006we}
\bibfield{author}{\bibinfo{person}{Alex~A Freitas}.} \bibinfo{year}{2006}\natexlab{}.
\newblock \showarticletitle{Are we really discovering interesting knowledge from data}.
\newblock \bibinfo{journal}{\emph{Expert Update (the BCS-SGAI magazine)}} \bibinfo{volume}{9}, \bibinfo{number}{1} (\bibinfo{year}{2006}), \bibinfo{pages}{41--47}.
\newblock


\bibitem[Freitas(2014)]%
        {freitas2014comprehensible}
\bibfield{author}{\bibinfo{person}{Alex~A Freitas}.} \bibinfo{year}{2014}\natexlab{}.
\newblock \showarticletitle{Comprehensible classification models: a position paper}.
\newblock \bibinfo{journal}{\emph{ACM SIGKDD explorations newsletter}} \bibinfo{volume}{15}, \bibinfo{number}{1} (\bibinfo{year}{2014}), \bibinfo{pages}{1--10}.
\newblock


\bibitem[Gr{\"u}ner et~al\mbox{.}(2020)]%
        {gruner2020reconstructing}
\bibfield{author}{\bibinfo{person}{Maximilian Gr{\"u}ner}, \bibinfo{person}{Melissa Licciardello}, {and} \bibinfo{person}{Ankit Singla}.} \bibinfo{year}{2020}\natexlab{}.
\newblock \showarticletitle{Reconstructing proprietary video streaming algorithms}. In \bibinfo{booktitle}{\emph{2020 USENIX Annual Technical Conference (USENIX ATC 20)}}.
\newblock


\bibitem[Huang et~al\mbox{.}(2019)]%
        {huang2019comyco}
\bibfield{author}{\bibinfo{person}{Tianchi Huang}, \bibinfo{person}{Chao Zhou}, \bibinfo{person}{Rui-Xiao Zhang}, \bibinfo{person}{Chenglei Wu}, \bibinfo{person}{Xin Yao}, {and} \bibinfo{person}{Lifeng Sun}.} \bibinfo{year}{2019}\natexlab{}.
\newblock \showarticletitle{Comyco: Quality-aware adaptive video streaming via imitation learning}. In \bibinfo{booktitle}{\emph{Proceedings of the 27th ACM international conference on multimedia}}. \bibinfo{pages}{429--437}.
\newblock


\bibitem[Huang et~al\mbox{.}(2014)]%
        {bba}
\bibfield{author}{\bibinfo{person}{Te-Yuan Huang}, \bibinfo{person}{Ramesh Johari}, \bibinfo{person}{Nick McKeown}, \bibinfo{person}{Matthew Trunnell}, {and} \bibinfo{person}{Mark Watson}.} \bibinfo{year}{2014}\natexlab{}.
\newblock \showarticletitle{A buffer-based approach to rate adaptation: Evidence from a large video streaming service}. In \bibinfo{booktitle}{\emph{Proceedings of the 2014 ACM conference on SIGCOMM}}. \bibinfo{pages}{187--198}.
\newblock


\bibitem[Huang et~al\mbox{.}(2023)]%
        {huang2023recommender}
\bibfield{author}{\bibinfo{person}{Xu Huang}, \bibinfo{person}{Jianxun Lian}, \bibinfo{person}{Yuxuan Lei}, \bibinfo{person}{Jing Yao}, \bibinfo{person}{Defu Lian}, {and} \bibinfo{person}{Xing Xie}.} \bibinfo{year}{2023}\natexlab{}.
\newblock \showarticletitle{Recommender ai agent: Integrating large language models for interactive recommendations}.
\newblock \bibinfo{journal}{\emph{arXiv preprint arXiv:2308.16505}} (\bibinfo{year}{2023}).
\newblock


\bibitem[Hussain et~al\mbox{.}(2024)]%
        {hussain2024tutorial}
\bibfield{author}{\bibinfo{person}{Zak Hussain}, \bibinfo{person}{Marcel Binz}, \bibinfo{person}{Rui Mata}, {and} \bibinfo{person}{Dirk~U Wulff}.} \bibinfo{year}{2024}\natexlab{}.
\newblock \showarticletitle{A tutorial on open-source large language models for behavioral science}.
\newblock \bibinfo{journal}{\emph{Behavior Research Methods}} \bibinfo{volume}{56}, \bibinfo{number}{8} (\bibinfo{year}{2024}), \bibinfo{pages}{8214--8237}.
\newblock


\bibitem[Huysmans et~al\mbox{.}(2011)]%
        {huysmans2011comprehensibility}
\bibfield{author}{\bibinfo{person}{Johan Huysmans}, \bibinfo{person}{Karel Dejaeger}, \bibinfo{person}{Christophe Mues}, \bibinfo{person}{Jan Vanthienen}, {and} \bibinfo{person}{Bart Baesens}.} \bibinfo{year}{2011}\natexlab{}.
\newblock \showarticletitle{An empirical evaluation of the comprehensibility of decision table, tree and rule based predictive models}.
\newblock \bibinfo{journal}{\emph{Decision Support Systems}} \bibinfo{volume}{51}, \bibinfo{number}{1} (\bibinfo{year}{2011}), \bibinfo{pages}{141--154}.
\newblock


\bibitem[Jia et~al\mbox{.}(2023)]%
        {jia2023rdladder}
\bibfield{author}{\bibinfo{person}{Lianchen Jia}, \bibinfo{person}{Chao Zhou}, \bibinfo{person}{Tianchi Huang}, \bibinfo{person}{Chaoyang Li}, {and} \bibinfo{person}{Lifeng Sun}.} \bibinfo{year}{2023}\natexlab{}.
\newblock \showarticletitle{Rdladder: Resolution-duration ladder for vbr-encoded videos via imitation learning}. In \bibinfo{booktitle}{\emph{IEEE INFOCOM 2023-IEEE Conference on Computer Communications}}. IEEE, \bibinfo{pages}{1--10}.
\newblock


\bibitem[Jia et~al\mbox{.}(2024)]%
        {jia2024dancing}
\bibfield{author}{\bibinfo{person}{Lianchen Jia}, \bibinfo{person}{Chao Zhou}, \bibinfo{person}{Tianchi Huang}, \bibinfo{person}{Chaoyang Li}, {and} \bibinfo{person}{Lifeng Sun}.} \bibinfo{year}{2024}\natexlab{}.
\newblock \showarticletitle{Dancing with Shackles, Meet the Challenge of Industrial Adaptive Streaming via Offline Reinforcement Learning}. In \bibinfo{booktitle}{\emph{IEEE INFOCOM 2024-IEEE Conference on Computer Communications}}. IEEE, \bibinfo{pages}{2169--2178}.
\newblock


\bibitem[Kaddour et~al\mbox{.}(2023)]%
        {kaddour2023challenges}
\bibfield{author}{\bibinfo{person}{Jean Kaddour}, \bibinfo{person}{Joshua Harris}, \bibinfo{person}{Maximilian Mozes}, \bibinfo{person}{Herbie Bradley}, \bibinfo{person}{Roberta Raileanu}, {and} \bibinfo{person}{Robert McHardy}.} \bibinfo{year}{2023}\natexlab{}.
\newblock \showarticletitle{Challenges and applications of large language models}.
\newblock \bibinfo{journal}{\emph{arXiv preprint arXiv:2307.10169}} (\bibinfo{year}{2023}).
\newblock


\bibitem[Kan et~al\mbox{.}(2022)]%
        {kan2022merina}
\bibfield{author}{\bibinfo{person}{Nuowen Kan}, \bibinfo{person}{Yuankun Jiang}, \bibinfo{person}{Chenglin Li}, \bibinfo{person}{Wenrui Dai}, \bibinfo{person}{Junni Zou}, {and} \bibinfo{person}{Hongkai Xiong}.} \bibinfo{year}{2022}\natexlab{}.
\newblock \showarticletitle{Improving generalization for neural adaptive video streaming via meta reinforcement learning}. In \bibinfo{booktitle}{\emph{Proceedings of the 30th ACM international conference on multimedia}}. \bibinfo{pages}{3006--3016}.
\newblock


\bibitem[Kaplan et~al\mbox{.}(2020)]%
        {kaplan2020scaling}
\bibfield{author}{\bibinfo{person}{Jared Kaplan}, \bibinfo{person}{Sam McCandlish}, \bibinfo{person}{Tom Henighan}, \bibinfo{person}{Tom~B Brown}, \bibinfo{person}{Benjamin Chess}, \bibinfo{person}{Rewon Child}, \bibinfo{person}{Scott Gray}, \bibinfo{person}{Alec Radford}, \bibinfo{person}{Jeffrey Wu}, {and} \bibinfo{person}{Dario Amodei}.} \bibinfo{year}{2020}\natexlab{}.
\newblock \showarticletitle{Scaling laws for neural language models}.
\newblock \bibinfo{journal}{\emph{arXiv preprint arXiv:2001.08361}} (\bibinfo{year}{2020}).
\newblock


\bibitem[Khorov et~al\mbox{.}(2018)]%
        {wifi6}
\bibfield{author}{\bibinfo{person}{Evgeny Khorov}, \bibinfo{person}{Anton Kiryanov}, \bibinfo{person}{Andrey Lyakhov}, {and} \bibinfo{person}{Giuseppe Bianchi}.} \bibinfo{year}{2018}\natexlab{}.
\newblock \showarticletitle{A tutorial on IEEE 802.11 ax high efficiency WLANs}.
\newblock \bibinfo{journal}{\emph{IEEE Communications Surveys \& Tutorials}} \bibinfo{volume}{21}, \bibinfo{number}{1} (\bibinfo{year}{2018}), \bibinfo{pages}{197--216}.
\newblock


\bibitem[Liang et~al\mbox{.}(2023)]%
        {liang2023encouraging}
\bibfield{author}{\bibinfo{person}{Tian Liang}, \bibinfo{person}{Zhiwei He}, \bibinfo{person}{Wenxiang Jiao}, \bibinfo{person}{Xing Wang}, \bibinfo{person}{Yan Wang}, \bibinfo{person}{Rui Wang}, \bibinfo{person}{Yujiu Yang}, \bibinfo{person}{Shuming Shi}, {and} \bibinfo{person}{Zhaopeng Tu}.} \bibinfo{year}{2023}\natexlab{}.
\newblock \showarticletitle{Encouraging divergent thinking in large language models through multi-agent debate}.
\newblock \bibinfo{journal}{\emph{arXiv preprint arXiv:2305.19118}} (\bibinfo{year}{2023}).
\newblock


\bibitem[Lin et~al\mbox{.}(2020)]%
        {lin2020gosdt}
\bibfield{author}{\bibinfo{person}{Jimmy Lin}, \bibinfo{person}{Chudi Zhong}, \bibinfo{person}{Diane Hu}, \bibinfo{person}{Cynthia Rudin}, {and} \bibinfo{person}{Margo Seltzer}.} \bibinfo{year}{2020}\natexlab{}.
\newblock \showarticletitle{Generalized and scalable optimal sparse decision trees}. In \bibinfo{booktitle}{\emph{International Conference on Machine Learning}}. PMLR, \bibinfo{pages}{6150--6160}.
\newblock


\bibitem[Loh(2011)]%
        {loh2011cart}
\bibfield{author}{\bibinfo{person}{Wei-Yin Loh}.} \bibinfo{year}{2011}\natexlab{}.
\newblock \showarticletitle{Classification and regression trees}.
\newblock \bibinfo{journal}{\emph{Wiley interdisciplinary reviews: data mining and knowledge discovery}} \bibinfo{volume}{1}, \bibinfo{number}{1} (\bibinfo{year}{2011}), \bibinfo{pages}{14--23}.
\newblock


\bibitem[Mao et~al\mbox{.}(2017)]%
        {mao2017pensieve}
\bibfield{author}{\bibinfo{person}{Hongzi Mao}, \bibinfo{person}{Ravi Netravali}, {and} \bibinfo{person}{Mohammad Alizadeh}.} \bibinfo{year}{2017}\natexlab{}.
\newblock \showarticletitle{Neural adaptive video streaming with pensieve}. In \bibinfo{booktitle}{\emph{Proceedings of the Conference of the ACM Special Interest Group on Data Communication}}. \bibinfo{pages}{197--210}.
\newblock


\bibitem[Martens et~al\mbox{.}(2007)]%
        {martens2007comprehensible}
\bibfield{author}{\bibinfo{person}{David Martens}, \bibinfo{person}{Bart Baesens}, \bibinfo{person}{Tony Van~Gestel}, {and} \bibinfo{person}{Jan Vanthienen}.} \bibinfo{year}{2007}\natexlab{}.
\newblock \showarticletitle{Comprehensible credit scoring models using rule extraction from support vector machines}.
\newblock \bibinfo{journal}{\emph{European journal of operational research}} \bibinfo{volume}{183}, \bibinfo{number}{3} (\bibinfo{year}{2007}), \bibinfo{pages}{1466--1476}.
\newblock


\bibitem[Martens et~al\mbox{.}(2011)]%
        {martens2011comprehensibilit}
\bibfield{author}{\bibinfo{person}{David Martens}, \bibinfo{person}{Jan Vanthienen}, \bibinfo{person}{Wouter Verbeke}, {and} \bibinfo{person}{Bart Baesens}.} \bibinfo{year}{2011}\natexlab{}.
\newblock \showarticletitle{Performance of classification models from a user perspective}.
\newblock \bibinfo{journal}{\emph{Decision Support Systems}} \bibinfo{volume}{51}, \bibinfo{number}{4} (\bibinfo{year}{2011}), \bibinfo{pages}{782--793}.
\newblock


\bibitem[McTavish et~al\mbox{.}(2022)]%
        {mctavish2022gosdt-guess}
\bibfield{author}{\bibinfo{person}{Hayden McTavish}, \bibinfo{person}{Chudi Zhong}, \bibinfo{person}{Reto Achermann}, \bibinfo{person}{Ilias Karimalis}, \bibinfo{person}{Jacques Chen}, \bibinfo{person}{Cynthia Rudin}, {and} \bibinfo{person}{Margo Seltzer}.} \bibinfo{year}{2022}\natexlab{}.
\newblock \showarticletitle{Fast sparse decision tree optimization via reference ensembles}. In \bibinfo{booktitle}{\emph{Proceedings of the AAAI conference on artificial intelligence}}, Vol.~\bibinfo{volume}{36}. \bibinfo{pages}{9604--9613}.
\newblock


\bibitem[Meng et~al\mbox{.}(2019)]%
        {meng2019pitree}
\bibfield{author}{\bibinfo{person}{Zili Meng}, \bibinfo{person}{Jing Chen}, \bibinfo{person}{Yaning Guo}, \bibinfo{person}{Chen Sun}, \bibinfo{person}{Hongxin Hu}, {and} \bibinfo{person}{Mingwei Xu}.} \bibinfo{year}{2019}\natexlab{}.
\newblock \showarticletitle{Pitree: Practical implementation of abr algorithms using decision trees}. In \bibinfo{booktitle}{\emph{Proceedings of the 27th ACM International Conference on Multimedia}}. \bibinfo{pages}{2431--2439}.
\newblock


\bibitem[Nakano et~al\mbox{.}(2021)]%
        {nakano2021webgpt}
\bibfield{author}{\bibinfo{person}{Reiichiro Nakano}, \bibinfo{person}{Jacob Hilton}, \bibinfo{person}{Suchir Balaji}, \bibinfo{person}{Jeff Wu}, \bibinfo{person}{Long Ouyang}, \bibinfo{person}{Christina Kim}, \bibinfo{person}{Christopher Hesse}, \bibinfo{person}{Shantanu Jain}, \bibinfo{person}{Vineet Kosaraju}, \bibinfo{person}{William Saunders}, {et~al\mbox{.}}} \bibinfo{year}{2021}\natexlab{}.
\newblock \showarticletitle{Webgpt: Browser-assisted question-answering with human feedback}.
\newblock \bibinfo{journal}{\emph{arXiv preprint arXiv:2112.09332}} (\bibinfo{year}{2021}).
\newblock


\bibitem[Narayanan et~al\mbox{.}(2020)]%
        {narayanan2020lumos5g}
\bibfield{author}{\bibinfo{person}{Arvind Narayanan}, \bibinfo{person}{Eman Ramadan}, \bibinfo{person}{Rishabh Mehta}, \bibinfo{person}{Xinyue Hu}, \bibinfo{person}{Qingxu Liu}, \bibinfo{person}{Rostand~AK Fezeu}, \bibinfo{person}{Udhaya~Kumar Dayalan}, \bibinfo{person}{Saurabh Verma}, \bibinfo{person}{Peiqi Ji}, \bibinfo{person}{Tao Li}, {et~al\mbox{.}}} \bibinfo{year}{2020}\natexlab{}.
\newblock \showarticletitle{Lumos5G: Mapping and predicting commercial mmWave 5G throughput}. In \bibinfo{booktitle}{\emph{Proceedings of the ACM Internet Measurement Conference}}. \bibinfo{pages}{176--193}.
\newblock


\bibitem[Narayanan et~al\mbox{.}(2021)]%
        {narayanan20215g}
\bibfield{author}{\bibinfo{person}{Arvind Narayanan}, \bibinfo{person}{Xumiao Zhang}, \bibinfo{person}{Ruiyang Zhu}, \bibinfo{person}{Ahmad Hassan}, \bibinfo{person}{Shuowei Jin}, \bibinfo{person}{Xiao Zhu}, \bibinfo{person}{Xiaoxuan Zhang}, \bibinfo{person}{Denis Rybkin}, \bibinfo{person}{Zhengxuan Yang}, \bibinfo{person}{Zhuoqing~Morley Mao}, {et~al\mbox{.}}} \bibinfo{year}{2021}\natexlab{}.
\newblock \showarticletitle{A variegated look at 5G in the wild: performance, power, and QoE implications}. In \bibinfo{booktitle}{\emph{Proceedings of the 2021 ACM SIGCOMM 2021 Conference}}. \bibinfo{pages}{610--625}.
\newblock


\bibitem[Networks(2025)]%
        {sandvine25}
\bibfield{author}{\bibinfo{person}{AppLogic Networks}.} \bibinfo{year}{2025}\natexlab{}.
\newblock \bibinfo{title}{global-internet-phenomena-report-2025}.
\newblock \bibinfo{howpublished}{https://www.applogicnetworks.com/phenomena}.
\newblock
\newblock
\shownote{[Online; accessed 30-May-2025]}.


\bibitem[OpenAI(2025)]%
        {GPT4o}
\bibfield{author}{\bibinfo{person}{OpenAI}.} \bibinfo{year}{2025}\natexlab{}.
\newblock \bibinfo{booktitle}{\emph{GPT4o}}.
\newblock
\urldef\tempurl%
\url{https://openai.com/index/hello-gpt-4o/}
\showURL{%
\tempurl}


\bibitem[Ouyang et~al\mbox{.}(2022)]%
        {ouyang2022instructgpt}
\bibfield{author}{\bibinfo{person}{Long Ouyang}, \bibinfo{person}{Jeffrey Wu}, \bibinfo{person}{Xu Jiang}, \bibinfo{person}{Diogo Almeida}, \bibinfo{person}{Carroll Wainwright}, \bibinfo{person}{Pamela Mishkin}, \bibinfo{person}{Chong Zhang}, \bibinfo{person}{Sandhini Agarwal}, \bibinfo{person}{Katarina Slama}, \bibinfo{person}{Alex Ray}, {et~al\mbox{.}}} \bibinfo{year}{2022}\natexlab{}.
\newblock \showarticletitle{Training language models to follow instructions with human feedback}.
\newblock \bibinfo{journal}{\emph{Advances in neural information processing systems}}  \bibinfo{volume}{35} (\bibinfo{year}{2022}), \bibinfo{pages}{27730--27744}.
\newblock


\bibitem[Park et~al\mbox{.}(2023)]%
        {park2023generative}
\bibfield{author}{\bibinfo{person}{Joon~Sung Park}, \bibinfo{person}{Joseph O'Brien}, \bibinfo{person}{Carrie~Jun Cai}, \bibinfo{person}{Meredith~Ringel Morris}, \bibinfo{person}{Percy Liang}, {and} \bibinfo{person}{Michael~S Bernstein}.} \bibinfo{year}{2023}\natexlab{}.
\newblock \showarticletitle{Generative agents: Interactive simulacra of human behavior}. In \bibinfo{booktitle}{\emph{Proceedings of the 36th Annual ACM Symposium on User Interface Software and Technology}}. \bibinfo{pages}{1--22}.
\newblock


\bibitem[Pazzani(2000)]%
        {pazzani2000knowledge}
\bibfield{author}{\bibinfo{person}{Michael~J Pazzani}.} \bibinfo{year}{2000}\natexlab{}.
\newblock \showarticletitle{Knowledge discovery from data?}
\newblock \bibinfo{journal}{\emph{IEEE intelligent systems and their applications}} \bibinfo{volume}{15}, \bibinfo{number}{2} (\bibinfo{year}{2000}), \bibinfo{pages}{10--12}.
\newblock


\bibitem[Pires and Simon(2015)]%
        {pires2015ugc}
\bibfield{author}{\bibinfo{person}{Karine Pires} {and} \bibinfo{person}{Gwendal Simon}.} \bibinfo{year}{2015}\natexlab{}.
\newblock \showarticletitle{YouTube live and Twitch: a tour of user-generated live streaming systems}. In \bibinfo{booktitle}{\emph{Proceedings of the 6th ACM multimedia systems conference}}. \bibinfo{pages}{225--230}.
\newblock


\bibitem[Raffel et~al\mbox{.}(2020)]%
        {raffel2020t5}
\bibfield{author}{\bibinfo{person}{Colin Raffel}, \bibinfo{person}{Noam Shazeer}, \bibinfo{person}{Adam Roberts}, \bibinfo{person}{Katherine Lee}, \bibinfo{person}{Sharan Narang}, \bibinfo{person}{Michael Matena}, \bibinfo{person}{Yanqi Zhou}, \bibinfo{person}{Wei Li}, {and} \bibinfo{person}{Peter~J Liu}.} \bibinfo{year}{2020}\natexlab{}.
\newblock \showarticletitle{Exploring the limits of transfer learning with a unified text-to-text transformer}.
\newblock \bibinfo{journal}{\emph{Journal of machine learning research}} \bibinfo{volume}{21}, \bibinfo{number}{140} (\bibinfo{year}{2020}), \bibinfo{pages}{1--67}.
\newblock


\bibitem[Ren et~al\mbox{.}(2024)]%
        {ren2024bases}
\bibfield{author}{\bibinfo{person}{Ruiyang Ren}, \bibinfo{person}{Peng Qiu}, \bibinfo{person}{Yingqi Qu}, \bibinfo{person}{Jing Liu}, \bibinfo{person}{Wayne~Xin Zhao}, \bibinfo{person}{Hua Wu}, \bibinfo{person}{Ji-Rong Wen}, {and} \bibinfo{person}{Haifeng Wang}.} \bibinfo{year}{2024}\natexlab{}.
\newblock \showarticletitle{BASES: Large-scale Web Search User Simulation with Large Language Model based Agents}.
\newblock \bibinfo{journal}{\emph{arXiv preprint arXiv:2402.17505}} (\bibinfo{year}{2024}).
\newblock


\bibitem[Report(2016)]%
        {fcc}
\bibfield{author}{\bibinfo{person}{Measuring Fixed~Broadband Report}.} \bibinfo{year}{2016}\natexlab{}.
\newblock \bibinfo{title}{Raw Data Measuring Broadband America 2016}.
\newblock \bibinfo{howpublished}{https://www.fcc.gov/reports-research/reports/measuring-broadband-america/raw-data-measuring-broadband-america-2016}.
\newblock
\newblock
\shownote{[Online; accessed 19-July-2016]}.


\bibitem[Riiser et~al\mbox{.}(2013)]%
        {riiser2013norway}
\bibfield{author}{\bibinfo{person}{Haakon Riiser}, \bibinfo{person}{Paul Vigmostad}, \bibinfo{person}{Carsten Griwodz}, {and} \bibinfo{person}{P{\aa}l Halvorsen}.} \bibinfo{year}{2013}\natexlab{}.
\newblock \showarticletitle{Commute path bandwidth traces from 3G networks: Analysis and applications}. In \bibinfo{booktitle}{\emph{Proceedings of the 4th ACM Multimedia Systems Conference}}. \bibinfo{pages}{114--118}.
\newblock


\bibitem[sandvine(2024)]%
        {sandvine24}
\bibfield{author}{\bibinfo{person}{sandvine}.} \bibinfo{year}{2024}\natexlab{}.
\newblock \bibinfo{title}{global-internet-phenomena-report-2024}.
\newblock \bibinfo{howpublished}{https://www.sandvine.com/global-internet-phenomena-report-2024}.
\newblock
\newblock
\shownote{[Online; accessed 30-May-2024]}.


\bibitem[Sani et~al\mbox{.}(2017)]%
        {sani2017abrsurvey}
\bibfield{author}{\bibinfo{person}{Yusuf Sani}, \bibinfo{person}{Andreas Mauthe}, {and} \bibinfo{person}{Christopher Edwards}.} \bibinfo{year}{2017}\natexlab{}.
\newblock \showarticletitle{Adaptive bitrate selection: A survey}.
\newblock \bibinfo{journal}{\emph{IEEE Communications Surveys \& Tutorials}} \bibinfo{volume}{19}, \bibinfo{number}{4} (\bibinfo{year}{2017}), \bibinfo{pages}{2985--3014}.
\newblock


\bibitem[Semenova et~al\mbox{.}(2022)]%
        {semenova2022Rashomon1}
\bibfield{author}{\bibinfo{person}{Lesia Semenova}, \bibinfo{person}{Cynthia Rudin}, {and} \bibinfo{person}{Ronald Parr}.} \bibinfo{year}{2022}\natexlab{}.
\newblock \showarticletitle{On the existence of simpler machine learning models}. In \bibinfo{booktitle}{\emph{Proceedings of the 2022 ACM Conference on Fairness, Accountability, and Transparency}}. \bibinfo{pages}{1827--1858}.
\newblock


\bibitem[Spiteri et~al\mbox{.}(2020)]%
        {spiteri2020bola}
\bibfield{author}{\bibinfo{person}{Kevin Spiteri}, \bibinfo{person}{Rahul Urgaonkar}, {and} \bibinfo{person}{Ramesh~K Sitaraman}.} \bibinfo{year}{2020}\natexlab{}.
\newblock \showarticletitle{BOLA: Near-optimal bitrate adaptation for online videos}.
\newblock \bibinfo{journal}{\emph{IEEE/ACM transactions on networking}} \bibinfo{volume}{28}, \bibinfo{number}{4} (\bibinfo{year}{2020}), \bibinfo{pages}{1698--1711}.
\newblock


\bibitem[Wang et~al\mbox{.}(2022)]%
        {wang2022consistency}
\bibfield{author}{\bibinfo{person}{Xuezhi Wang}, \bibinfo{person}{Jason Wei}, \bibinfo{person}{Dale Schuurmans}, \bibinfo{person}{Quoc Le}, \bibinfo{person}{Ed Chi}, \bibinfo{person}{Sharan Narang}, \bibinfo{person}{Aakanksha Chowdhery}, {and} \bibinfo{person}{Denny Zhou}.} \bibinfo{year}{2022}\natexlab{}.
\newblock \showarticletitle{Self-consistency improves chain of thought reasoning in language models}.
\newblock \bibinfo{journal}{\emph{arXiv preprint arXiv:2203.11171}} (\bibinfo{year}{2022}).
\newblock


\bibitem[Wu et~al\mbox{.}(2024)]%
        {wu2024netllm}
\bibfield{author}{\bibinfo{person}{Duo Wu}, \bibinfo{person}{Xianda Wang}, \bibinfo{person}{Yaqi Qiao}, \bibinfo{person}{Zhi Wang}, \bibinfo{person}{Junchen Jiang}, \bibinfo{person}{Shuguang Cui}, {and} \bibinfo{person}{Fangxin Wang}.} \bibinfo{year}{2024}\natexlab{}.
\newblock \showarticletitle{Netllm: Adapting large language models for networking}. In \bibinfo{booktitle}{\emph{Proceedings of the ACM SIGCOMM 2024 Conference}}. \bibinfo{pages}{661--678}.
\newblock


\bibitem[Xia et~al\mbox{.}(2022)]%
        {xia2022genet}
\bibfield{author}{\bibinfo{person}{Zhengxu Xia}, \bibinfo{person}{Yajie Zhou}, \bibinfo{person}{Francis~Y Yan}, {and} \bibinfo{person}{Junchen Jiang}.} \bibinfo{year}{2022}\natexlab{}.
\newblock \showarticletitle{Genet: automatic curriculum generation for learning adaptation in networking}. In \bibinfo{booktitle}{\emph{Proceedings of the ACM SIGCOMM 2022 Conference}}. \bibinfo{pages}{397--413}.
\newblock


\bibitem[Xin et~al\mbox{.}(2022)]%
        {xin2022treefarms}
\bibfield{author}{\bibinfo{person}{Rui Xin}, \bibinfo{person}{Chudi Zhong}, \bibinfo{person}{Zhi Chen}, \bibinfo{person}{Takuya Takagi}, \bibinfo{person}{Margo Seltzer}, {and} \bibinfo{person}{Cynthia Rudin}.} \bibinfo{year}{2022}\natexlab{}.
\newblock \showarticletitle{Exploring the whole rashomon set of sparse decision trees}.
\newblock \bibinfo{journal}{\emph{Advances in neural information processing systems}}  \bibinfo{volume}{35} (\bibinfo{year}{2022}), \bibinfo{pages}{14071--14084}.
\newblock


\bibitem[Yan et~al\mbox{.}(2020)]%
        {yan2020fugu}
\bibfield{author}{\bibinfo{person}{Francis~Y Yan}, \bibinfo{person}{Hudson Ayers}, \bibinfo{person}{Chenzhi Zhu}, \bibinfo{person}{Sadjad Fouladi}, \bibinfo{person}{James Hong}, \bibinfo{person}{Keyi Zhang}, \bibinfo{person}{Philip~Alexander Levis}, {and} \bibinfo{person}{Keith Winstein}.} \bibinfo{year}{2020}\natexlab{}.
\newblock \showarticletitle{Learning in situ: a randomized experiment in video streaming.}. In \bibinfo{booktitle}{\emph{NSDI}}, Vol.~\bibinfo{volume}{20}. \bibinfo{pages}{495--511}.
\newblock


\bibitem[Yin et~al\mbox{.}(2015)]%
        {yin2015mpc}
\bibfield{author}{\bibinfo{person}{Xiaoqi Yin}, \bibinfo{person}{Abhishek Jindal}, \bibinfo{person}{Vyas Sekar}, {and} \bibinfo{person}{Bruno Sinopoli}.} \bibinfo{year}{2015}\natexlab{}.
\newblock \showarticletitle{A control-theoretic approach for dynamic adaptive video streaming over HTTP}. In \bibinfo{booktitle}{\emph{Proceedings of the 2015 ACM Conference on Special Interest Group on Data Communication}}. \bibinfo{pages}{325--338}.
\newblock


\bibitem[Zhao et~al\mbox{.}(2023b)]%
        {zhao2023llmsurvey}
\bibfield{author}{\bibinfo{person}{Wayne~Xin Zhao}, \bibinfo{person}{Kun Zhou}, \bibinfo{person}{Junyi Li}, \bibinfo{person}{Tianyi Tang}, \bibinfo{person}{Xiaolei Wang}, \bibinfo{person}{Yupeng Hou}, \bibinfo{person}{Yingqian Min}, \bibinfo{person}{Beichen Zhang}, \bibinfo{person}{Junjie Zhang}, \bibinfo{person}{Zican Dong}, {et~al\mbox{.}}} \bibinfo{year}{2023}\natexlab{b}.
\newblock \showarticletitle{A survey of large language models}.
\newblock \bibinfo{journal}{\emph{arXiv preprint arXiv:2303.18223}} (\bibinfo{year}{2023}).
\newblock


\bibitem[Zhao et~al\mbox{.}(2023a)]%
        {zhao2023recommender}
\bibfield{author}{\bibinfo{person}{Zihuai Zhao}, \bibinfo{person}{Wenqi Fan}, \bibinfo{person}{Jiatong Li}, \bibinfo{person}{Yunqing Liu}, \bibinfo{person}{Xiaowei Mei}, \bibinfo{person}{Yiqi Wang}, \bibinfo{person}{Zhen Wen}, \bibinfo{person}{Fei Wang}, \bibinfo{person}{Xiangyu Zhao}, \bibinfo{person}{Jiliang Tang}, {et~al\mbox{.}}} \bibinfo{year}{2023}\natexlab{a}.
\newblock \showarticletitle{Recommender systems in the era of large language models (llms)}.
\newblock \bibinfo{journal}{\emph{arXiv preprint arXiv:2307.02046}} (\bibinfo{year}{2023}).
\newblock


\end{thebibliography}
